\documentclass{article}

\usepackage{epsfig}
\usepackage{graphics}
\usepackage{graphicx}
\usepackage[centertags]{amsmath}
\usepackage{amsfonts}
\usepackage{amsthm}
\usepackage{amssymb}
\usepackage{url}
\usepackage{subfigure}
\usepackage{hyperref}

\begin{document}

\title{Cellular automaton supercolliders}

\author{Genaro J. Mart\'{\i}nez$^{1,2}$, Andrew Adamatzky$^1$ \\ Christopher R. Stephens$^2$, Alejandro F. Hoeflich$^2$}


\maketitle

\begin{centering}
$^1$ Unconventional Computing Centre, University of the West of England, Bristol BS16 1QY, United Kingdom. \\
\{genaro.martinez, andrew.adamatzky\}@uwe.ac.uk \\
$^2$ Instituto de Ciencias Nucleares and Centro de Ciencias de la Complejidad, Universidad Nacional Aut\'onoma de M\'exico, M\'exico. \\
\{stephens, frank\}@nucleares.unam.mx \\
\end{centering}

\begin{abstract}
\noindent
Gliders in one-dimensional cellular automata are compact groups of non-quiescent and non-ether patterns (ether represents a periodic background) translating along automaton lattice. They are cellular-automaton analogous of localizations or quasi-local collective excitations travelling in a spatially extended non-linear medium. They can be considered as binary strings or symbols travelling along a one-dimensional ring, interacting with each other and changing their states, or symbolic values, as a result of interactions. We analyse what types of interaction occur between gliders travelling on a cellular automaton `cyclotron' and build a catalog of the most common reactions. We demonstrate that collisions between gliders emulate the basic types of interaction that occur between localizations in non-linear media: fusion, elastic collision, and soliton-like collision. Computational outcomes of a swarm of gliders circling on a one-dimensional torus are analysed via implementation of cyclic tag systems. \\

\end{abstract}

\section{Introduction}

The era of unconventional computers --- implementations of computing schemes in physical, chemical and biological substrates, and interpretation of the behaviour of natural systems' in terms of computation -- has brought us a plethora of original prototypes of future and emerging computing architectures. Some examples of this new paradigm are polymer- and conductive foam based extended analog computers~\cite{mills_2008}, chemical reaction-diffusion computers~\cite{gorecki_2003,kn:ACA05}, aromatic molecular computers~\cite{bandyo_2010}, slime mould computers~\cite{nakagaki_2000maze}, micro-fluidic based computers~\cite{fuerstman_2003}, enzymatic logical circuits~\cite{katz_2010}, molecular arithmetical circuits~\cite{stojanovic_2003} to name but a few. Many of these novel computing devices work on the principle that information is represented by localised states of natural systems (phase or diffusive waves, propagating pseudopodia, electron density, conformation) which can then be represented by cellular automata or other discrete automata models (at least in principle)~\cite{kn:Ada01}. Most models of unconventional computers suffer, up to different degrees, from boundary problems. The systems are confined to some experimental arena, e.g. test tube or a Petri dish, or even be controlled and programmed by externally imposed geometrical constraints (e.g. channels in excitable chemical computers). It would be incredible useful to produce a model which is self-contained, is not dependent on external constraints and which can, subject to resources and energy supplied, function indefinitely. A particle ``super collider'' may be the best candidate for such a universal model of boundary free computation.

In his seminal paper ``Symbol super colliders''~\cite{kn:Toff02} Tommaso Toffoli envisaged far fetching analogies between physical implementations of particle super colliders, lattice gas and one-dimensional cellular automata. He suggested that the concept of symbol super collider -- where myriad of tokens run along intersecting rings and interact with each other to produce new tokens -- can be used in designing novel types of nature-inspired computing devices \cite{kn:FT01, kn:Toff98}. In present paper we develop Toffoli's ideas  further and provide computational implementations of particle `accelerator' or test rigs, implemented in elementary cellular automata. 

The paper is structured as follows. One-dimensional cellular automata, ahistoric and with memory, are introduced in Sect.~\ref{ca}. The concepts of supercollider and ballistic computing are presented in Sect.~\ref{collider}. Section~\ref{ballisticCollision} shows how essential elements of collision-based computers can be implemented via glider interactions in one-dimensional cellular automata, and computational capacities are presented in Sect.~\ref{routing}. Outcomes of the presented results and directions for further studies are discussed in Sect.~\ref{summary}.

\section{One-dimensional cellular automata}
\label{ca}

A cellular automaton (CA) is a quintuple $ \langle \Sigma,\varphi,\mu, c_0, L \rangle$  based on a one-dimensional lattice $L$ of cells, where each cell $x_i$, $i \in N$, takes a state from a finite alphabet $\Sigma$. A sequence $s \in \Sigma^n$ of $n$ cell-states represents a string or a global configuration $c$ on $\Sigma$. We write a set of finite configurations as $\Sigma^n$. Cells update their states via an evolution rule $\varphi: \Sigma^{\mu} \rightarrow \Sigma$, such that $\mu$ represents a cell neighbourhood that consists of a central cell and a number of neighbours connected locally.  There are $|\Sigma|^{\mu}$ different neighbourhoods and if $k=|\Sigma|$ then we have $k^{k^n}$ different evolution rules.

An evolution diagram for a CA is represented by a sequence of configurations $\{c_i\}$ generated by the global mapping $\Phi:\Sigma^n \rightarrow \Sigma^n$, where a global relation is given by as $\Phi(c^t) \rightarrow c^{t+1}$. $c_0$ is the initial configuration. Cell states of a configuration $c^t$ are updated simultaneously by the evolution rule as:

\begin{equation}
\varphi(x_{i-r}^t, \ldots, x_{i}^t, \ldots, x_{i+r}^t) \rightarrow x_i^{t+1}.
\end{equation}

\noindent  where $i$ indicates cell position and $r$ is the radius of neighbourhood $\mu$. Thus, the elementary basic CA class represents a system of order $(k=2,\ r=1)$ (in Wolfram's nomenclature \cite{kn:Wolf02}), the well-known ECA. To represent a specific evolution rule we write name of the rule in a decimal notation, e.g. $\varphi_{R110}$.

Conventional cellular automata are memoryless:  the new state of a cell depends on the neighbourhood configuration solely at the preceding time step of $\varphi$. CA with memory are an extension of ECA in such a way that every cell $x_i$ is allowed to remember its states during some fixed period of its evolution. CA with memory have been proposed originally by Alonso-Sanz~\cite{kn:AM03, kn:Alo06, kn:Alo09}.

We implement a memory function $\phi$ as follow:

\begin{equation}
\phi (x^{t-\tau}_{i}, \ldots, x^{t-1}_{i}, x^{t}_{i}) \rightarrow s_{i} ,
\end{equation}

\noindent where $\tau < t$ determines the degree of memory and each cell $s_{i} \in \Sigma$ is a state function of the series of states of the cell $x_i$ with memory up $\tau$. To execute the evolution we apply the original rule $\varphi(\ldots, s^{t}_{i-1}, s^{t}_{i}, s^{t}_{i+1}, \ldots) \rightarrow x^{t+1}_i$. Thus in CA with memory,  while the mapping $\varphi$ remains unaltered, historic memory of all past iterations is retained by featuring each cell as a summary of its past states from $\phi$. We can say that cells canalize memory to the map $\varphi$~\cite{kn:Alo09}.

Let us consider the memory function $\phi$ as a majority memory $\phi_{maj} \rightarrow s_{i}$, where in case of a tie given by $\Sigma_1 = \Sigma_0$ from $\phi$ we take the last value $x_i$. So $\phi_{maj}$ function represents the classic majority function for three values \cite{kn:Mins67} as follows:

\begin{equation*}
(a \wedge b) \vee (b \wedge c) \vee (c \wedge a)
\end{equation*}

\noindent that represents the cells $(x^{t-\tau}_{i}, \ldots, x^{t-1}_{i}, x^{t}_{i})$ and defines a temporal ring before to get the next global configuration $c$. Of course, this evaluation can be for any number of values. In this way, a number of functional memories may be used such as: minority, parity, alpha, etc. (see \cite{kn:Alo09}).

\begin{figure}[!tbp]
\centerline{\includegraphics[width=4.8in]{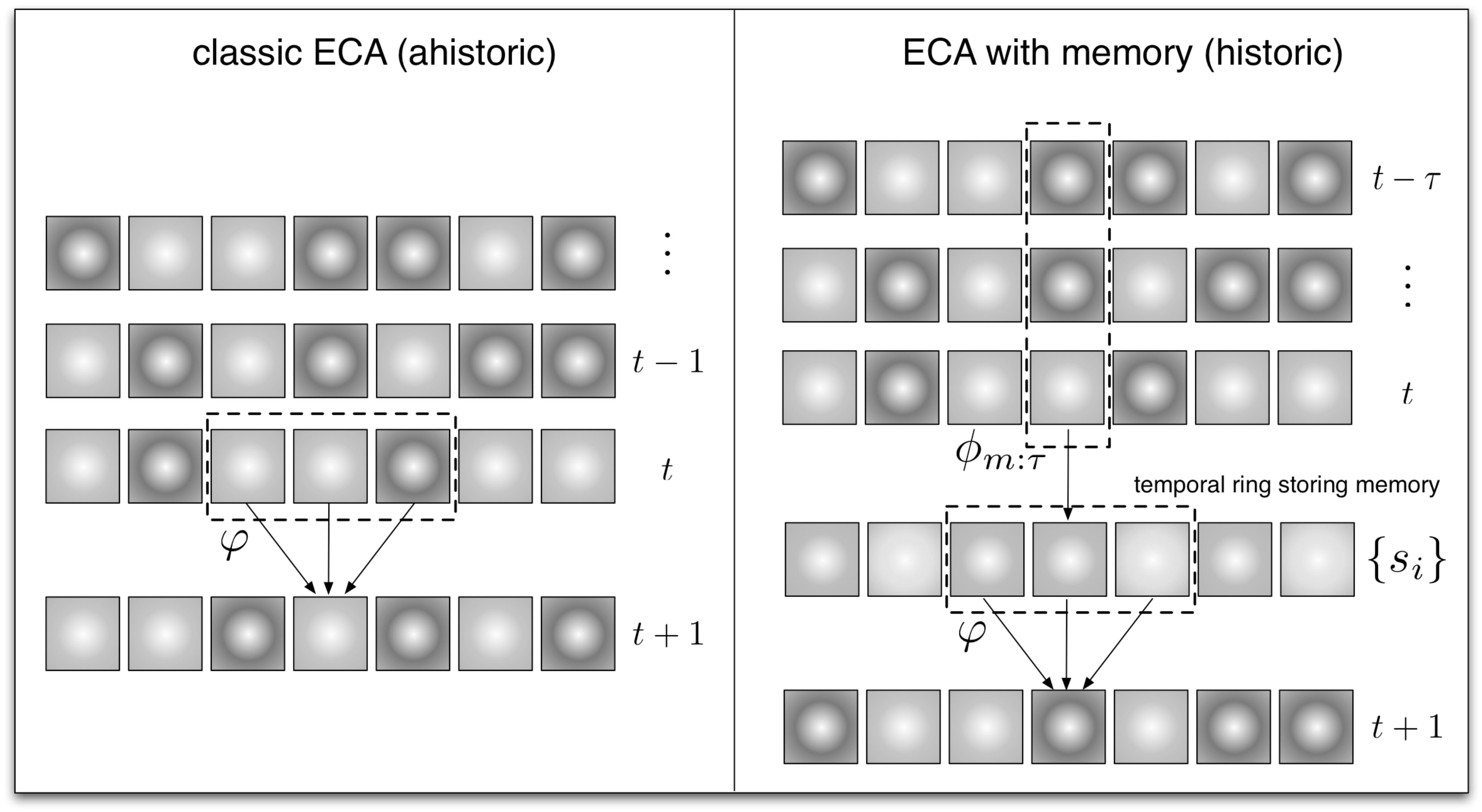}}
\caption{Illustration of cell-state transition in ECA (left) and ECA with memory (right).}
\label{memEvol}
\end{figure}

Evolution rules representation in ECA with memory as given in  \cite{kn:MAA10,kn:MAS10} is as follows:

\begin{equation}
\phi_{CARm:\tau}
\end{equation}

\noindent where $CAR$ is a decimal notation of a particular ECA rule and $m$ is a kind of memory used given with a specific value of $\tau$. Thus for example, the majority memory ($maj$) incorporated in ECA Rule 30 employing five step of a cell's history ($\tau=5$)  is denoted as $\phi_{R30maj:5}$. The memory is a function of the CA itself, see schematic explanation in Fig.~\ref{memEvol}.

\section{Toffoli's supercollider}
\label{collider}

In the late 1970s Fredkin and Toffoli developed a concept of a general-purpose computation based on ballistic interactions between quanta of information that are represented by abstract particles~\cite{kn:Toff02}. The Boolean states of logical variables are represented by balls or atoms, which preserve their identity when they collide with each other. They came up with the idea of a billiard-ball model of computation, with underpinning mechanics of elastically colliding balls and mirrors reflecting the balls' trajectories. Later Margolus developed a special class of CA which implement the billiard-ball model. Margolus' partitioned CA  exhibited computational universality because they simulated Fredkin gate via collision of soft spheres~\cite{kn:Marg03}.

\begin{figure}[th]
\centering
\subfigure[]{\scalebox{0.45}{\includegraphics{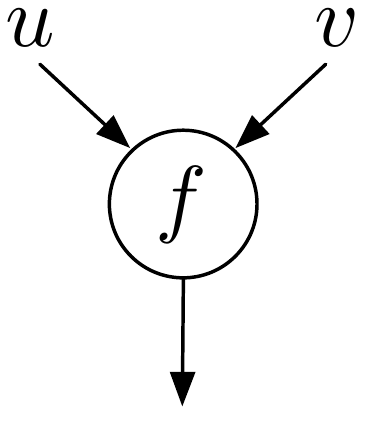}}} \hspace{1cm}
\subfigure[]{\scalebox{0.45}{\includegraphics{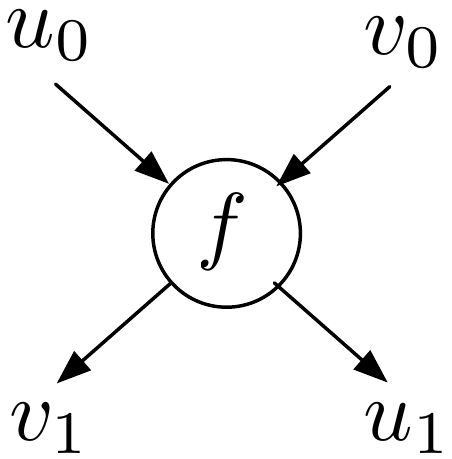}}} \hspace{1cm}
\subfigure[]{\scalebox{0.45}{\includegraphics{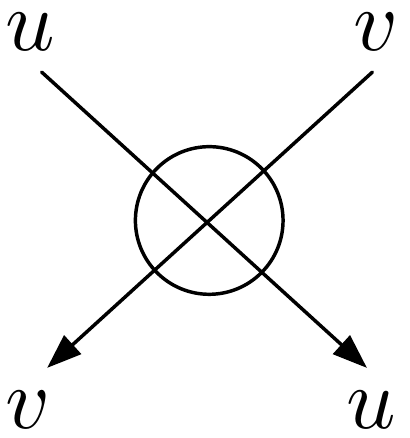}}}
\hspace{1cm}
\subfigure[]{\scalebox{0.45}{\includegraphics{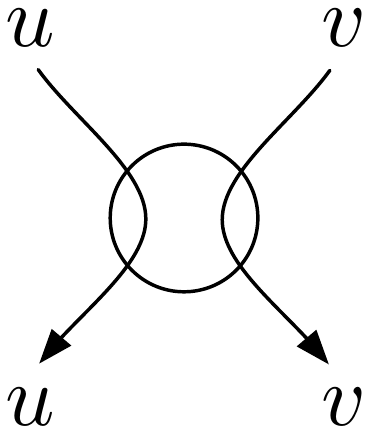}}}
\caption{Basic schemes of ballistic collision between localizations representing logical values of the Boolean variables $u$ and $v$.}
\label{particles}
\end{figure}

The following basic functions with two input arguments $u$ and $v$ can be expressed via collision between two localizations:

\begin{enumerate}
\item $f(u,v) = c$, fusion (Fig.~\ref{particles}a)
\item $f(u,v) = u+v$, interaction and subsequent change of state (Fig.~\ref{particles}b)
\item $f_i(u,v) \mapsto (u,v)$ identity, solitonic collision (Fig.~\ref{particles}c);
\item $f_r(u,v) \mapsto (v,u)$ reflection, elastic collision (Fig.~\ref{particles}d);
\end{enumerate}

To map Toffoli's supercollider~\cite{kn:Toff02} onto a one-dimensional CA we use the notion of an idealized particle $p \in \Sigma^+$ (without energy and potential energy). The particle $p$ is represented by a binary string of cell states.

\begin{figure}[th]
\centering
\subfigure[]{\scalebox{0.5}{\includegraphics{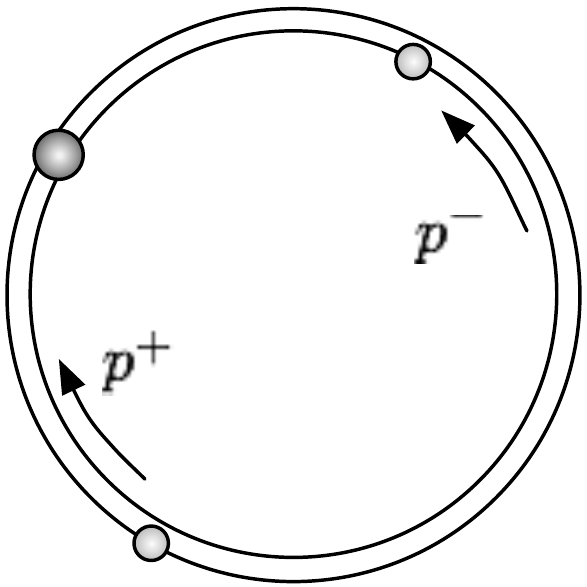}}} \hspace{1cm}
\subfigure[]{\scalebox{0.5}{\includegraphics{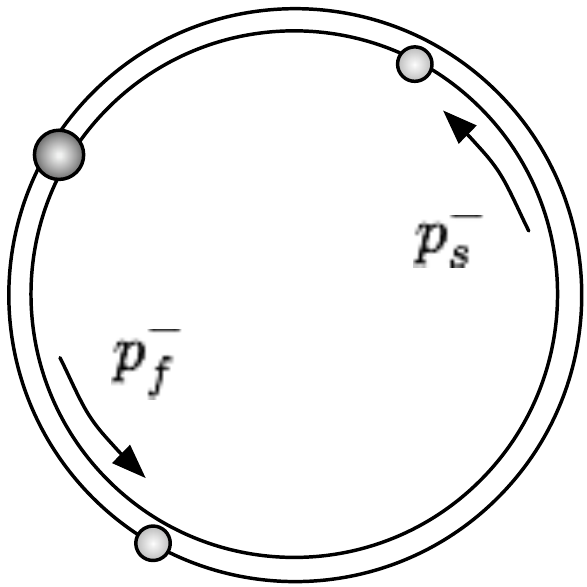}}} \hspace{1cm}
\subfigure[]{\scalebox{0.5}{\includegraphics{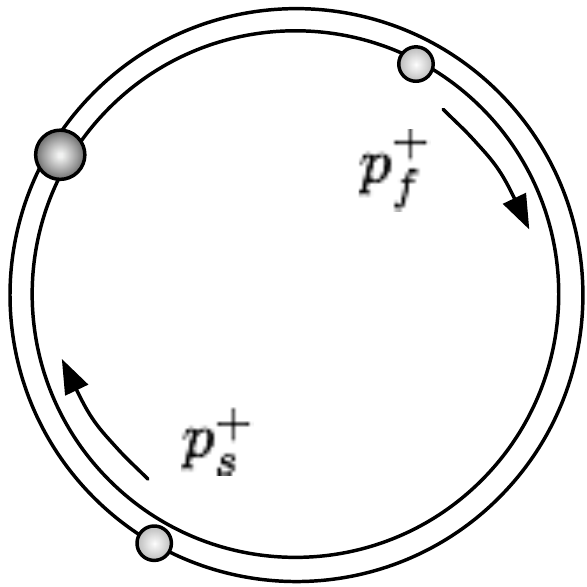}}}
\caption{Representation of abstract particles in a one-dimensional CA ring.}
\label{beamRouting}
\end{figure}

Figure~\ref{beamRouting} shows two typical scenarios where particles $p_f$ and $p_S$ travel in a CA cyclotron. The first scenario (Fig.~\ref{beamRouting}a) shows two particles travelling in opposite directions which then collide. Their collision site is shown by a dark circle in (Fig.~\ref{beamRouting}a). The second scenario demonstrates a typical beam routing where a fast particle $p_f$ eventually catches up with a slow particle $p_s$ at a collision site (Fig.~\ref{beamRouting}b). If the particles collide like solitons~\cite{kn:JSS01}, then the faster particle $p_f$ simply overtakes the slower particle $p_s$ and continues its motion (Fig.~\ref{beamRouting}c).

\begin{figure}[th]
\centerline{\includegraphics[width=3in]{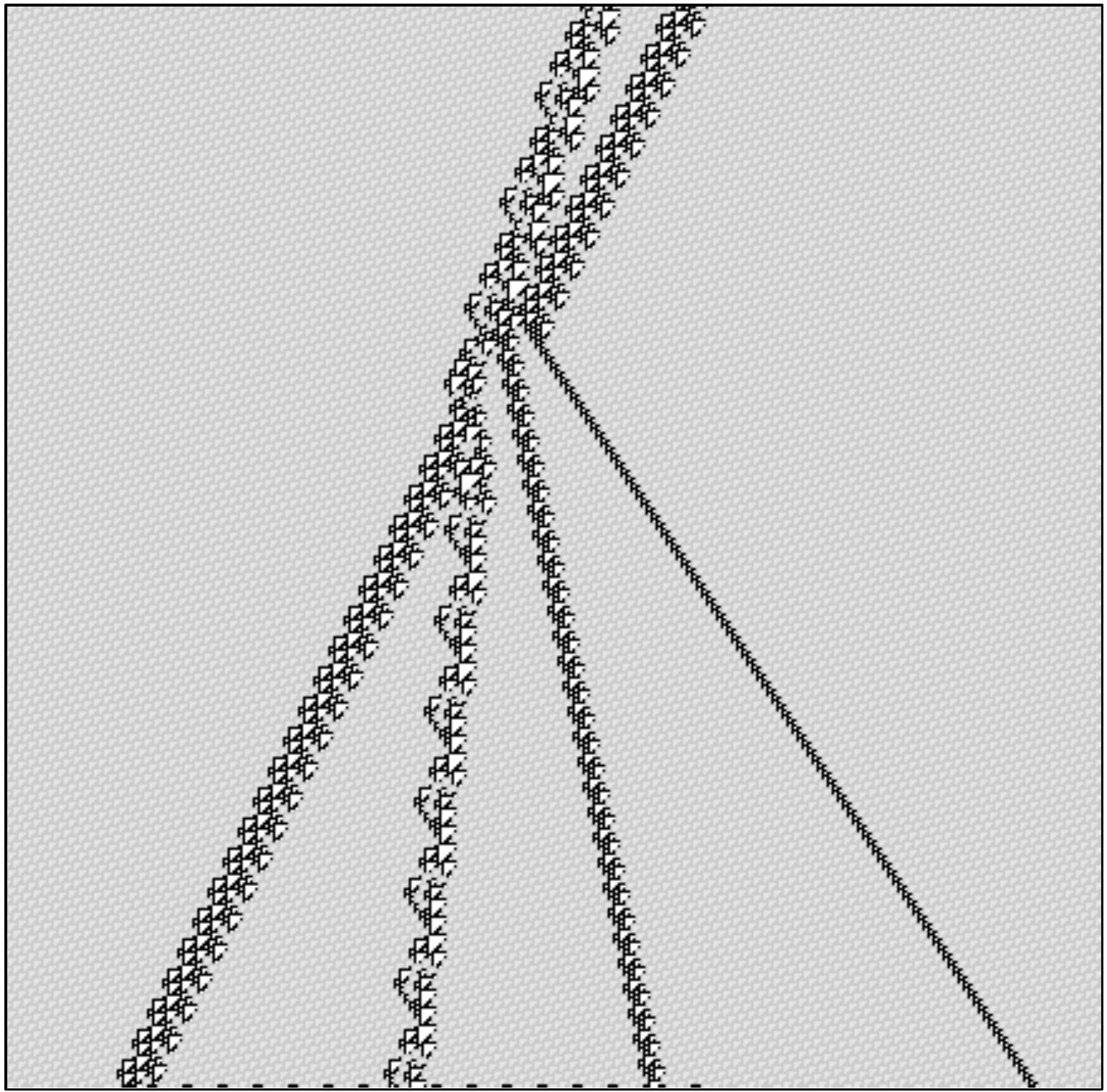}}
\caption{Example of a particle collision in $\varphi_{R110}$. Particle $p^-_{\bar{B}}$ collides with particle $p^-_{G}$ giving rise to three new particles --- $p^-_{F}$, $p^+_{D_2}$, and $p^+_{A^3}$ --- that are generated as a result of the collision. }
\label{splitParticleR110}
\end{figure}

Typically, we can find all types of particles manifest in CA gliders, including positive $p^+$, negative $p^-$, and neutral $p^0$ displacements \cite{kn:MMS06}, and also composite particles assembled from elementary localizations. Let us consider the case where a quiescent state is substituted by cells synchronized together as an ether (periodic background). This phenomenon is associated with ECA Rule 110 $\varphi_{R110}$.\footnote{Rule 110 repository \url{http://uncomp.uwe.ac.uk/genaro/Rule110.html}} Its evolution space is dominated by a number of particles emerging in various different orders, some of which are really quite complex constructions. Consequently, the number of collisions between particles is increased. Each particle has a period, displacement, velocity, mass, volume, and phase~\cite{kn:MMS06, kn:MMS08}.\footnote{A full description of particles in Rule 110 is available at \url{http://uncomp.uwe.ac.uk/genaro/rule110/glidersRule110.html}} Figure~\ref{splitParticleR110} displays a typical collision between two particles in $\varphi_{R110}$. As a result of the collision one particle is split into three different particles (for details please see \cite{kn:MM01}). The pre-collision positions of particles determines the outcomes of the collision.

\begin{figure}[th]
\centering
\subfigure[]{\scalebox{0.303}{\includegraphics{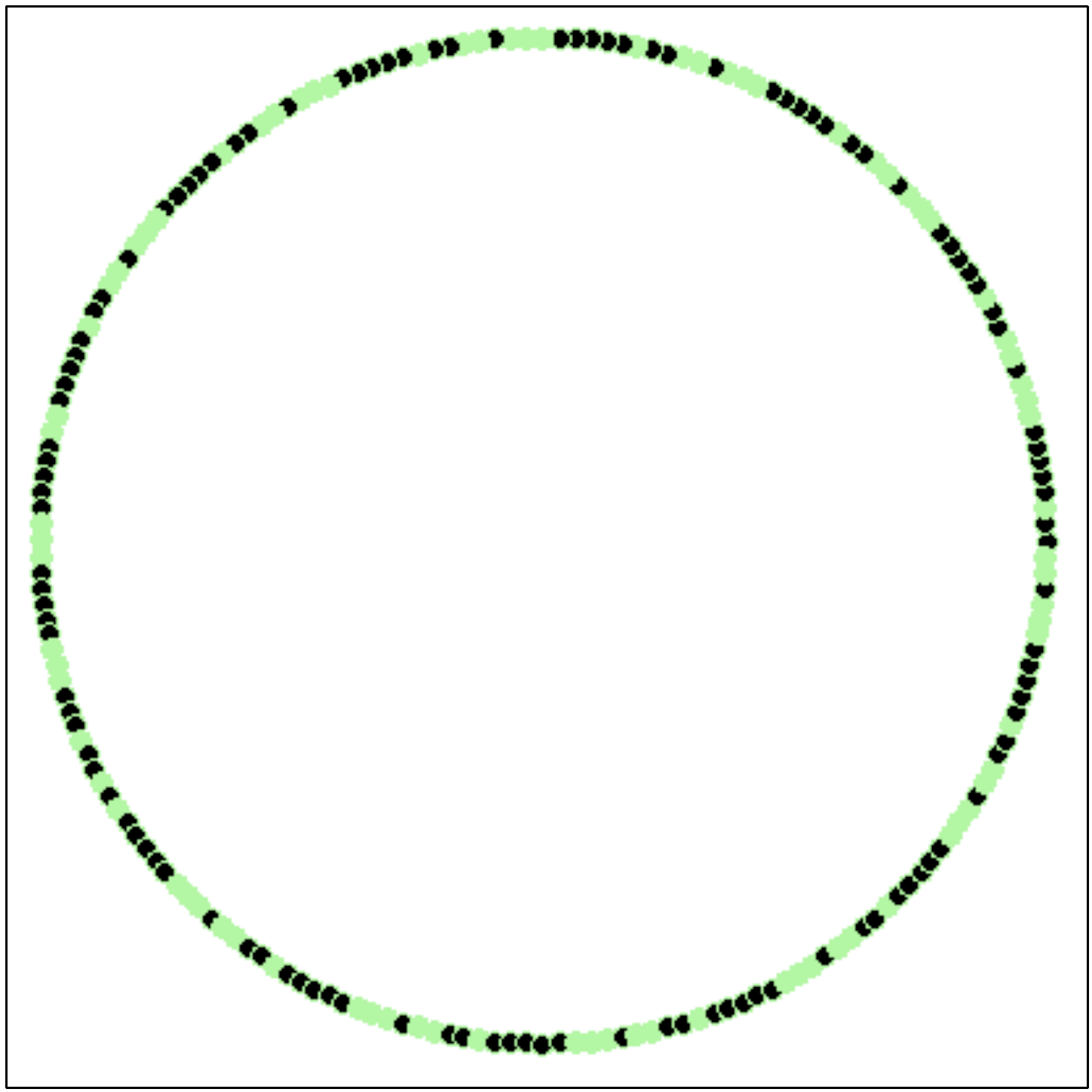}}} 
\subfigure[]{\scalebox{0.312}{\includegraphics{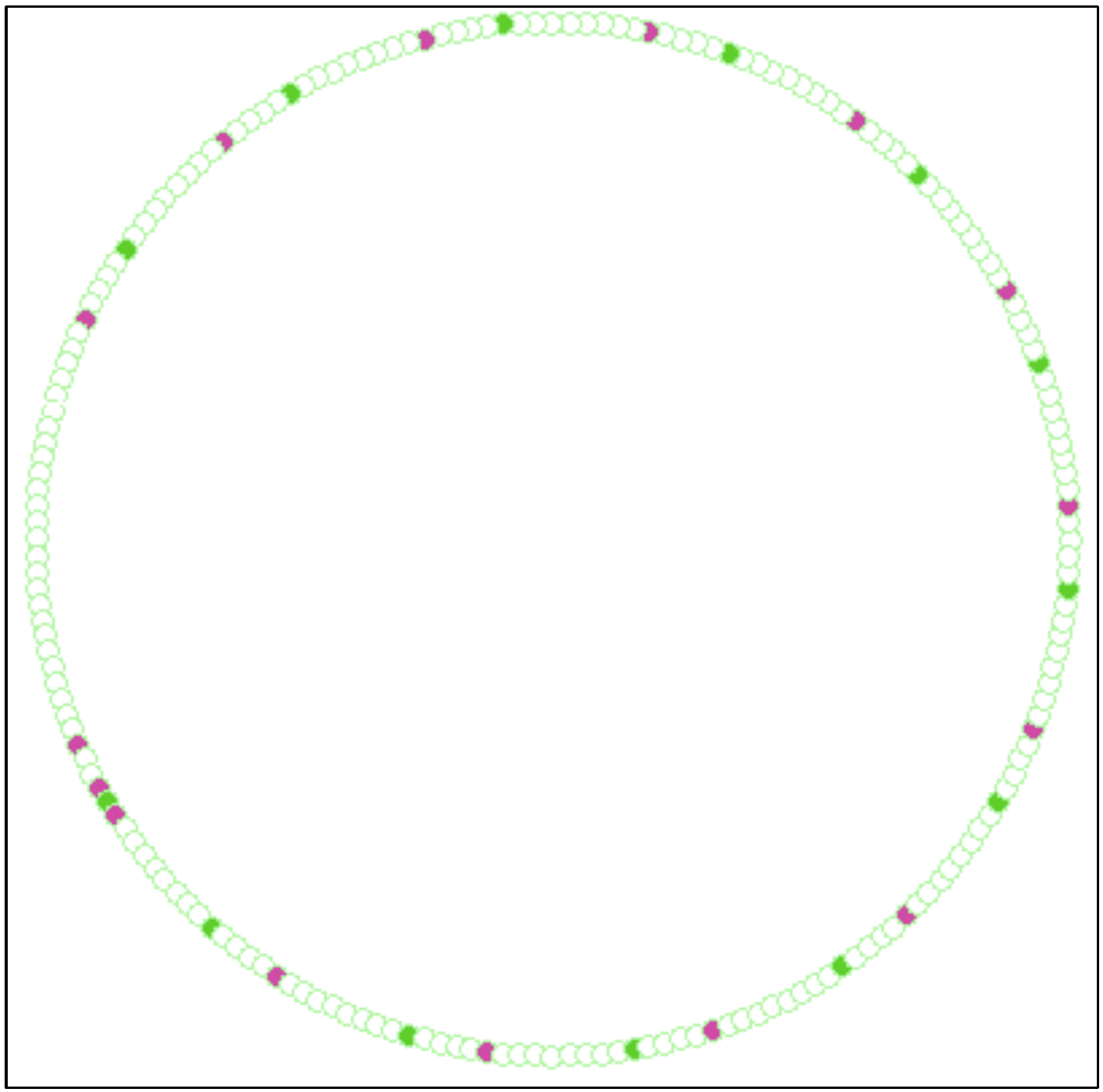}}}  
\subfigure[]{\scalebox{0.308}{\includegraphics{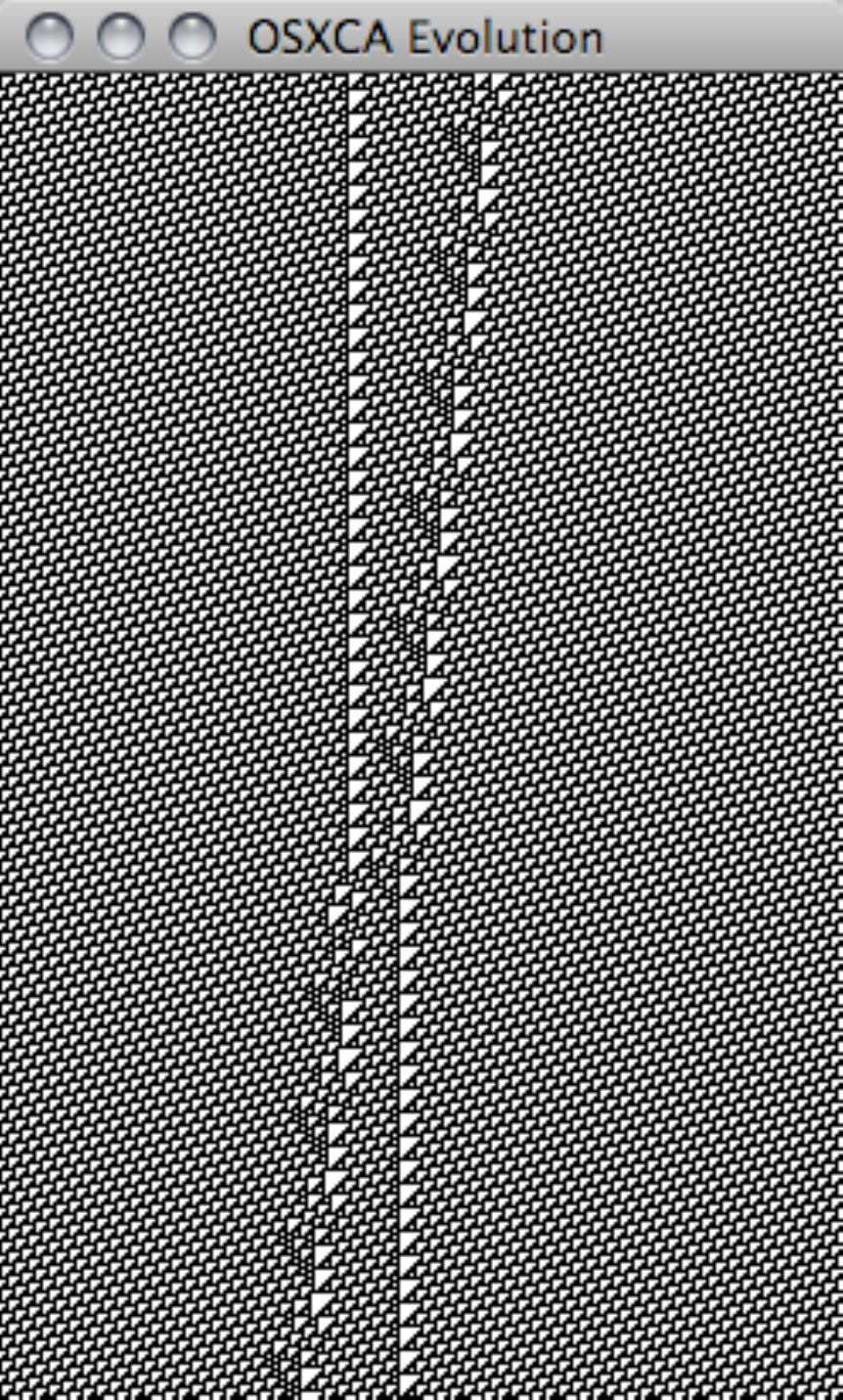}}} 
\caption{Example of a soliton-type interaction between particles in one-dimensional cellular automaton $\varphi_{R110}$: (a)--(b)~two steps of beam routing, (c)~exact configuration at the time of collision.}
\label{solitonR110}
\end{figure}

To represent particles on a given beam routing scheme (see Fig.~\ref{beamRouting}), we do not consider ether configuration in $\varphi_{R110}$ because this does not affect on collisions. Figure~\ref{solitonR110} displays a one-dimensional configuration where two particles collide repeatedly and interact as solitonic so that the identities of the particles are preserved in the collisions. A negative particle $p^-_F$ collides and overtakes a neutral particle $p^-_{C_1}$. Figure~\ref{solitonR110}a presents a whole set of cells in state 1 (dark points) where the ether configuration makes it impossible to distinguish the particles: $p^-_F$ and $p^-_{C_1}$. However, we can apply a filter and thereby select particles from their background ether (Fig.~\ref{solitonR110}b).\footnote{Ring evolution was simulated with DDLab available in \url{http://www.ddlab.org}.} Space-time configurations of a cellular automaton exhibiting a collision between particles $p^-_F$ and $p^-_{C_1}$ are shown in Fig.~\ref{solitonR110}c.

Filters selected in CA are a useful tool for understand ``hidden'' properties of CA. This tool was amply developed by Wuensche in the context of automatic classification of CA \cite{kn:Wue99}. The filters were derived from mechanical computation techniques \cite{kn:HC97}, pattern recognition \cite{kn:SHR05}, and analysis of cell-state frequencies \cite{kn:Wue99}. Thus, a filter is a sequence of cells that have a high frequency in the evolution space. Such $d$-dimensional string repeat periodically, coexisting with any complex structure without altering or disturbing the global dynamics.

\section{Ballistic collisions in cellular automata}
\label{ballisticCollision}

In this section we analyze examples of ballistic collisions between particles in ECA and ECA with memory. Let us start with ECA with memory governed by the rule $\phi_{R22maj:4}$~\cite{kn:MMA}. There are only two types of particles in this rule $\cal G$$_{\phi_{R22maj:4}} = \{g_{L},g_{R}\}$. Their properties are easy to infer: Both particles have a volume of a perfect square of $11 \times 11$ cells, a mass of 35 cells, and they translate 2 cells per 11 time steps (or iterations of CA evolution). The particle $g_L$ has a negative slope with a velocity of $-\frac{2}{11}$ and the particle $g_R$ has a positive slope with a velocity of $\frac{2}{11}$.

\begin{figure}
\centering
\subfigure[]{\scalebox{0.41}{\includegraphics{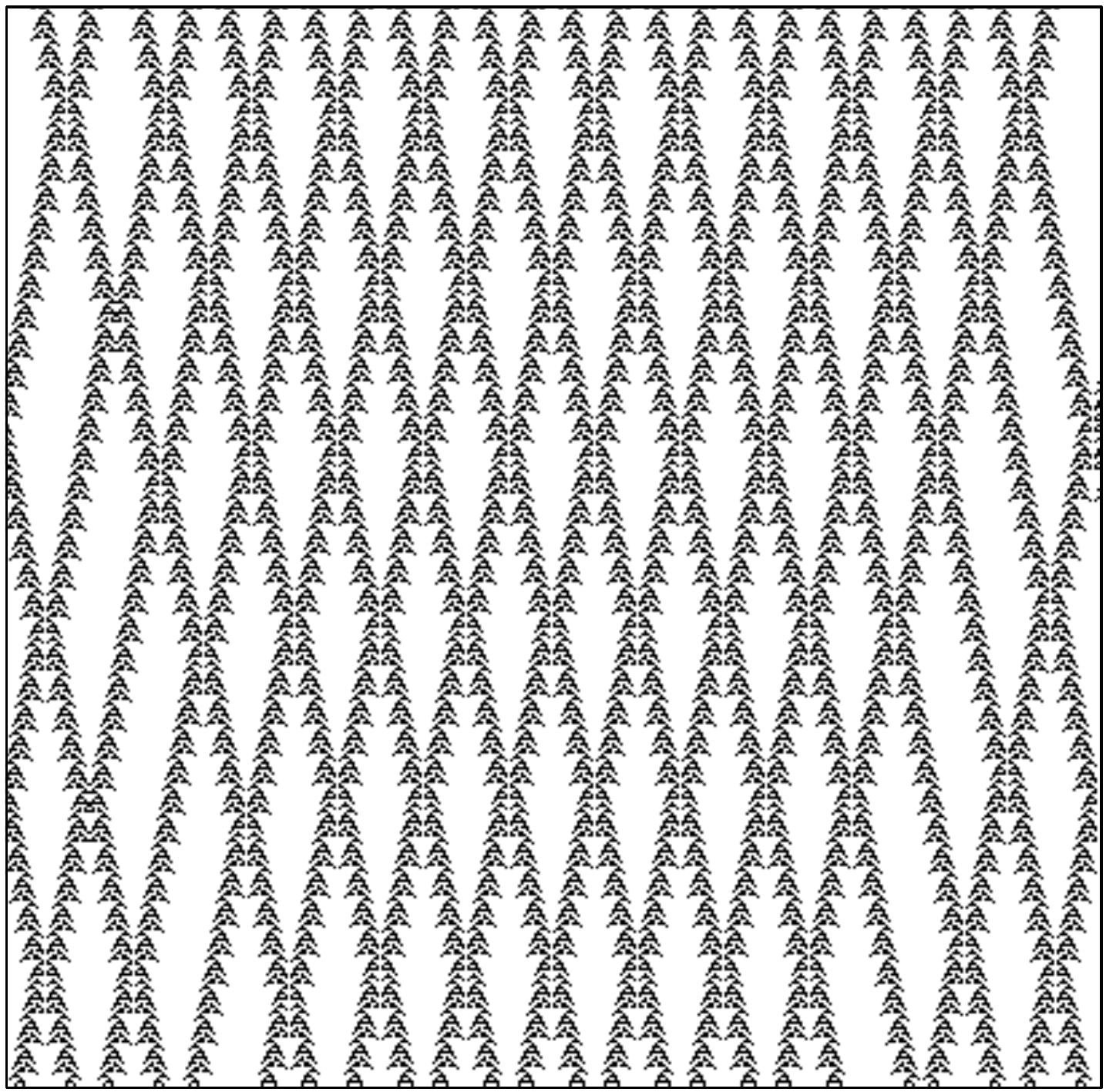}}} 
\subfigure[]{\scalebox{0.41}{\includegraphics{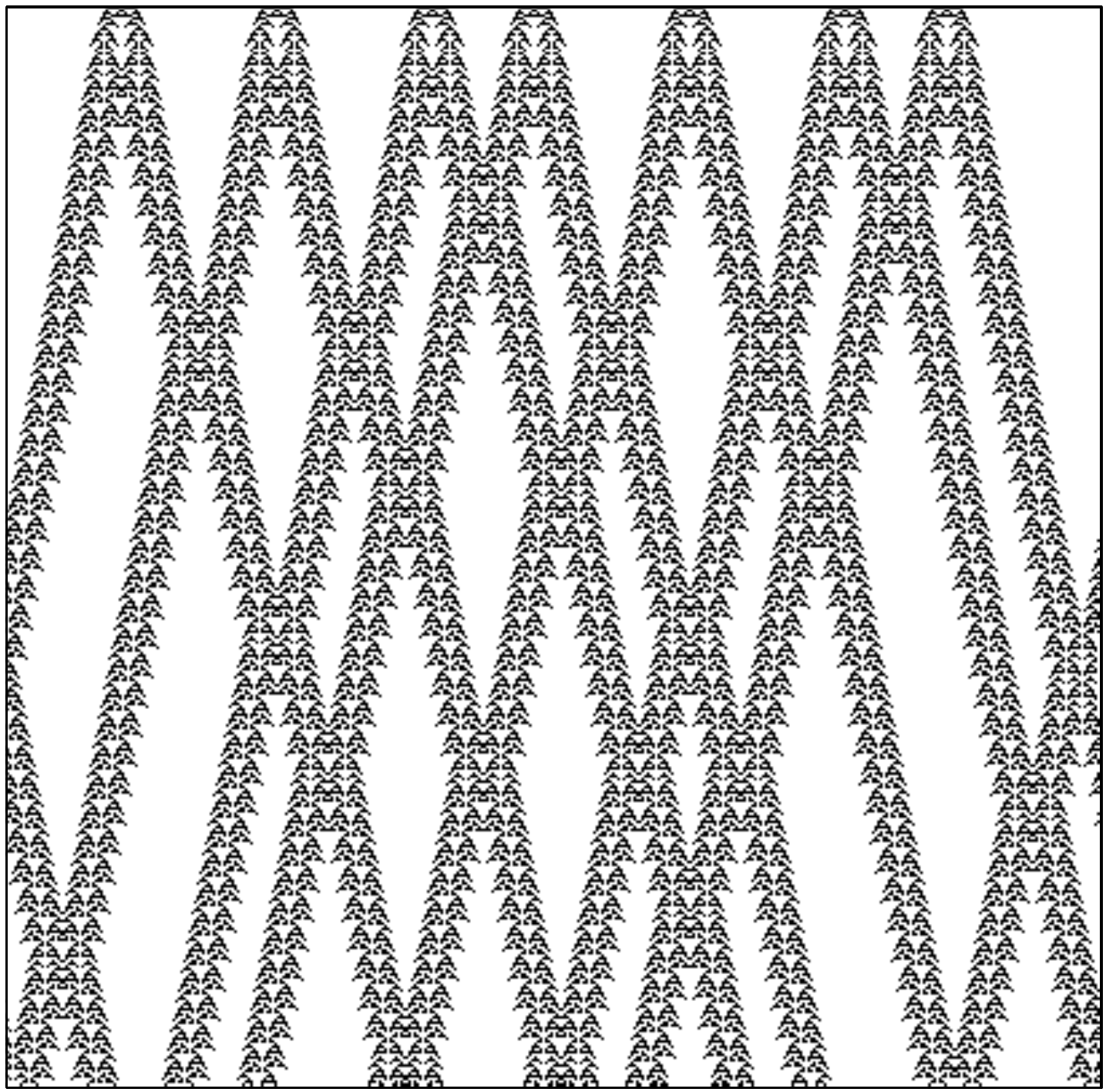}}}
\caption{Ballistic collisions in CA $\phi_{R22maj:4}$: (a)~identity or soliton collision, and (b)~reflections.}
\label{ballistic}
\end{figure}

The rule $\phi_{R22maj:4}$ supports two types of ballistic collisions: $f_i(u,v)$ and $f_r(u,v)$. Figure~\ref{ballistic}a shows how a number of soliton interactions can be synchronized as a identity collision $f_{i}(u,v)$, where both particles can cross their own trajectories. This result can be achieved by selecting a particular phase of each particle as encoded by its initial condition. Different initial conditions lead to different reactions~\cite{kn:MMA}. For example, in Fig.~\ref{ballistic}b we can see particles undergoing elastic collisions similarly to a lattice gas model~\cite{kn:Toff02}. This is a reflection collisions $f_{r}(u,v)$.

\begin{table}[th]
\centering
\small
\begin{tabular}{|c|c|}
\hline
particle & velocity \\
\hline \hline
$\overrightarrow{w}$ & 2/2 = 1 \\
\hline
$\overleftarrow{w}$ & -2/2 = -1 \\
\hline
$g_{o}$ & 0/4 = 0 \\
\hline
$g_{e}$ & 0/4 = 0 \\
\hline
gun &  0/32 = 0 \\
\hline
\end{tabular}
\caption{Properties of particles in rule $\varphi_{R54}$ \cite{kn:MAM06}.}
\label{tablaGlidersR54}
\end{table}

\begin{figure}
\centering
\subfigure[]{\scalebox{0.385}{\includegraphics{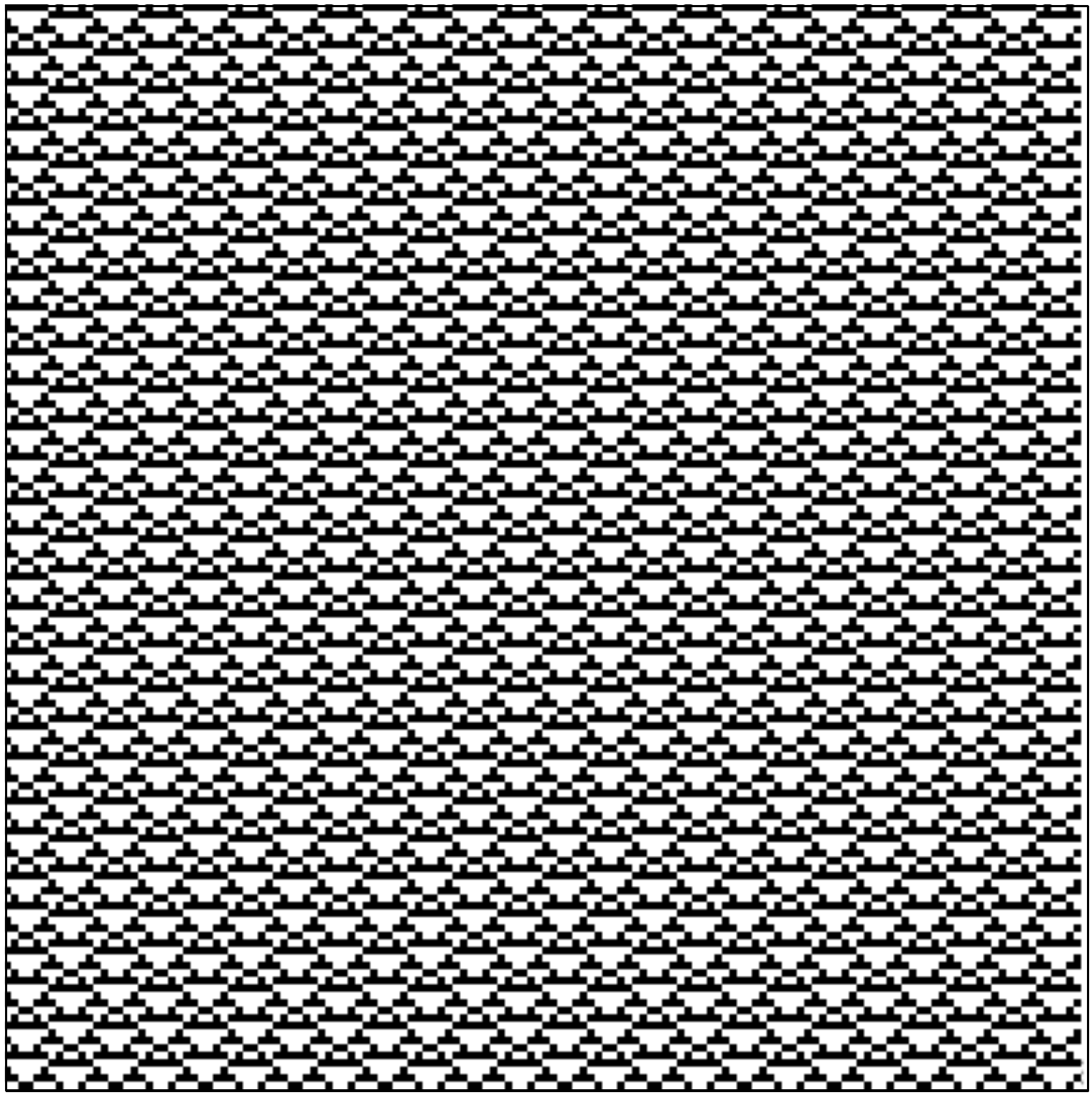}}} 
\subfigure[]{\scalebox{0.385}{\includegraphics{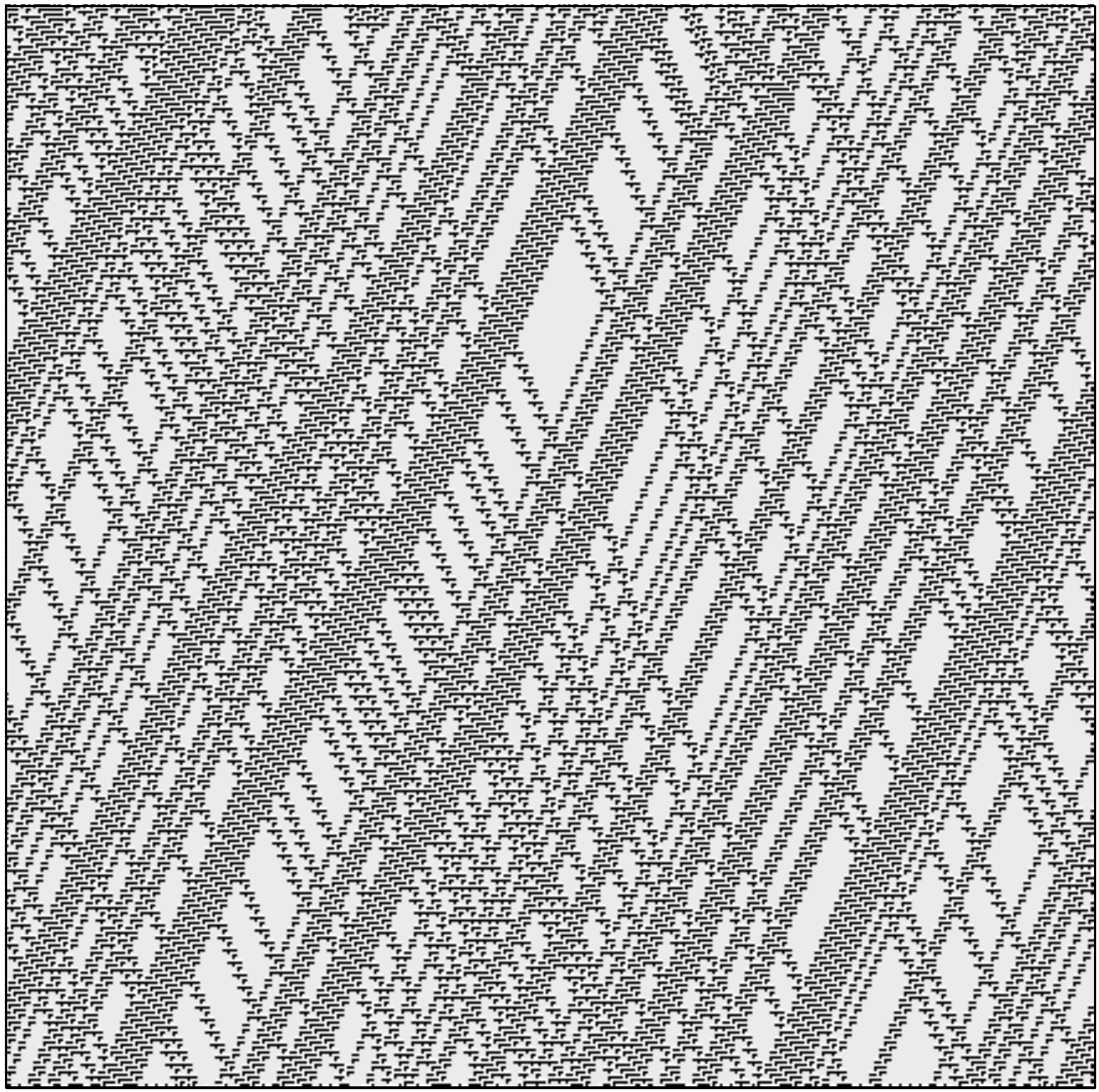}}} 
\caption{Ballistic collisions simulating the identity relation, or a solitonic reaction: (a)~with $\varphi_{R54}$ and (b)~with $\phi_{R9maj:4}$.}
\label{soliton}
\end{figure}

Figure~\ref{soliton} presents solitonic reactions (identity ballistic collision) for rules  $\varphi_{R54}$ (memoryless ECA) and $\phi_{R9maj:4}$ (ECA with memory). The rule $\varphi_{R54}$ has been studied in detail in~\cite{kn:BNR91, kn:HC97, kn:MAM06, kn:MAM08}.\footnote{Rule 54 repository in \url{http://uncomp.uwe.ac.uk/genaro/Rule54.html}} In representing particles in $\varphi_{R54}$ as $\cal G$$_{\varphi_{R54}} = \{ \overrightarrow{w}, \overleftarrow{w}, g_{o}, g_{e}, \mbox{gun}\}$, we follow Boccara's {\it et al.} notation~\cite{kn:BNR91}. Table~\ref{tablaGlidersR54} gives the relation between particles and velocity.\footnote{Velocity is calculated as the number of cells a particle displaces in one unit of discrete time.}. In this way, to produce a required reaction we must code a particular initial condition, where a $\overrightarrow{w}$ particle is ready to collide with a $\overleftarrow{w}$ particle and both particles collide with the same phase with a stationary  particle  $g_{e}$ (Fig.~\ref{soliton}a).

Figure~\ref{soliton}(b) shows an ECA with memory which preserves identity ballistic collision starting with any random initial condition. Thus, $\phi_{R9maj:4}$ evolves with two particles $\cal G$$_{\phi_{R9maj:4}} = \{ \overrightarrow{p}, \overleftarrow{p} \}$. The particles' properties are easy to calculate. The $\overrightarrow{p}$ particle has a volume of $5 \times 6$ cells, a mass of 12 cells, and moves 2 cells in 5 generations (positive slope). While $\overleftarrow{p}$ particle has a volume of $5 \times 3$ cells, a mass of 7 cells, and moves 2 cells in 5 generations (negative slope).

As the example above demonstrates, by using ECA and ECA with memory we can experimentally study a variety of ballistic collisions.

\section{Beam routings and computations}
\label{routing}

In this section we will exploit beam routing to produce some complex constructions that are based on particle-collisions. An additional effort is necessary to code initial conditions for every particle and recognize the most suitable phase for each particle in order to produce the desired reactions. Also, we will show how the beam routing can be used in design of computing based-collisions.

\begin{figure}
\centering
\subfigure[]{\scalebox{0.35}{\includegraphics{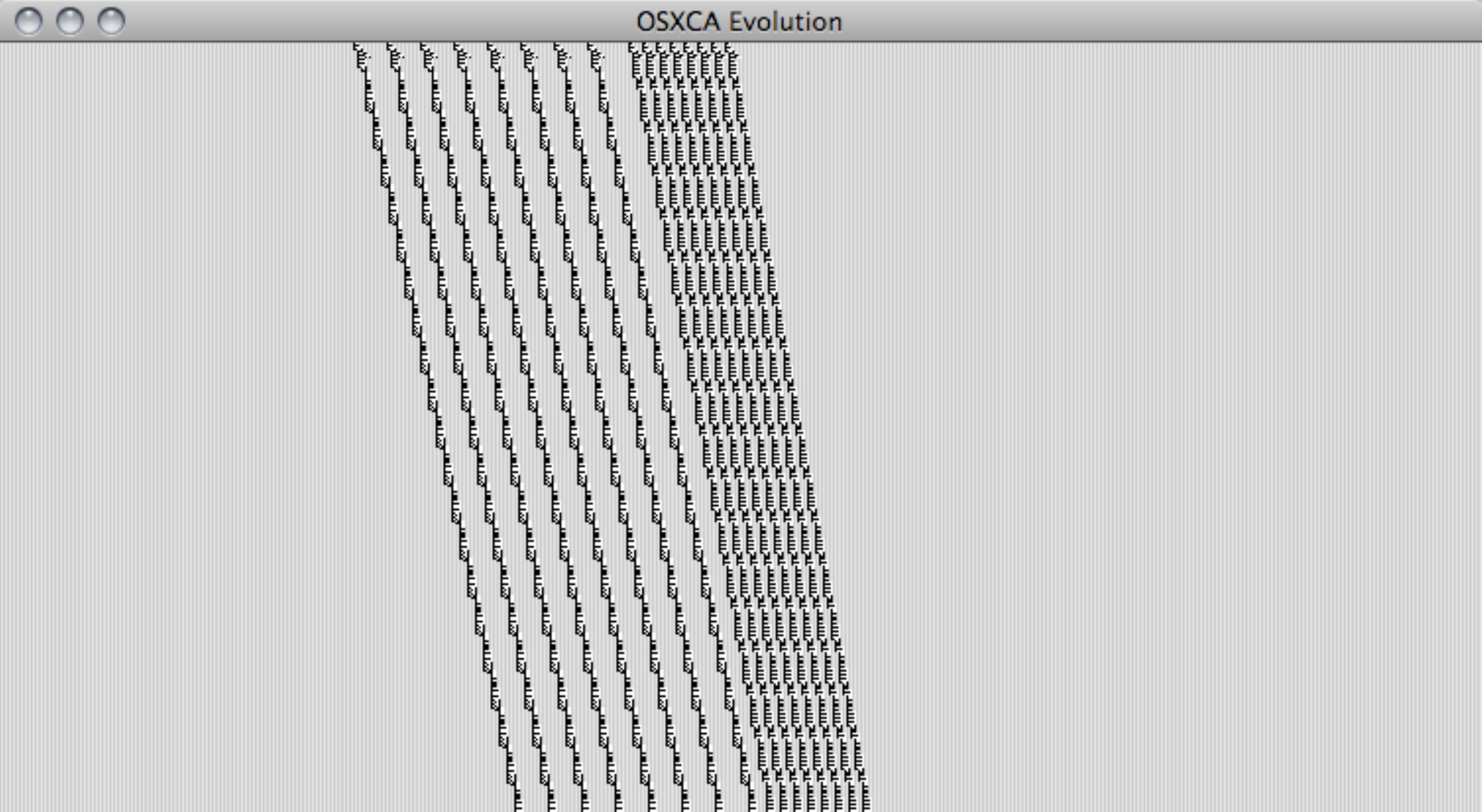}}} \hspace{0.7cm}
\subfigure[]{\scalebox{0.35}{\includegraphics{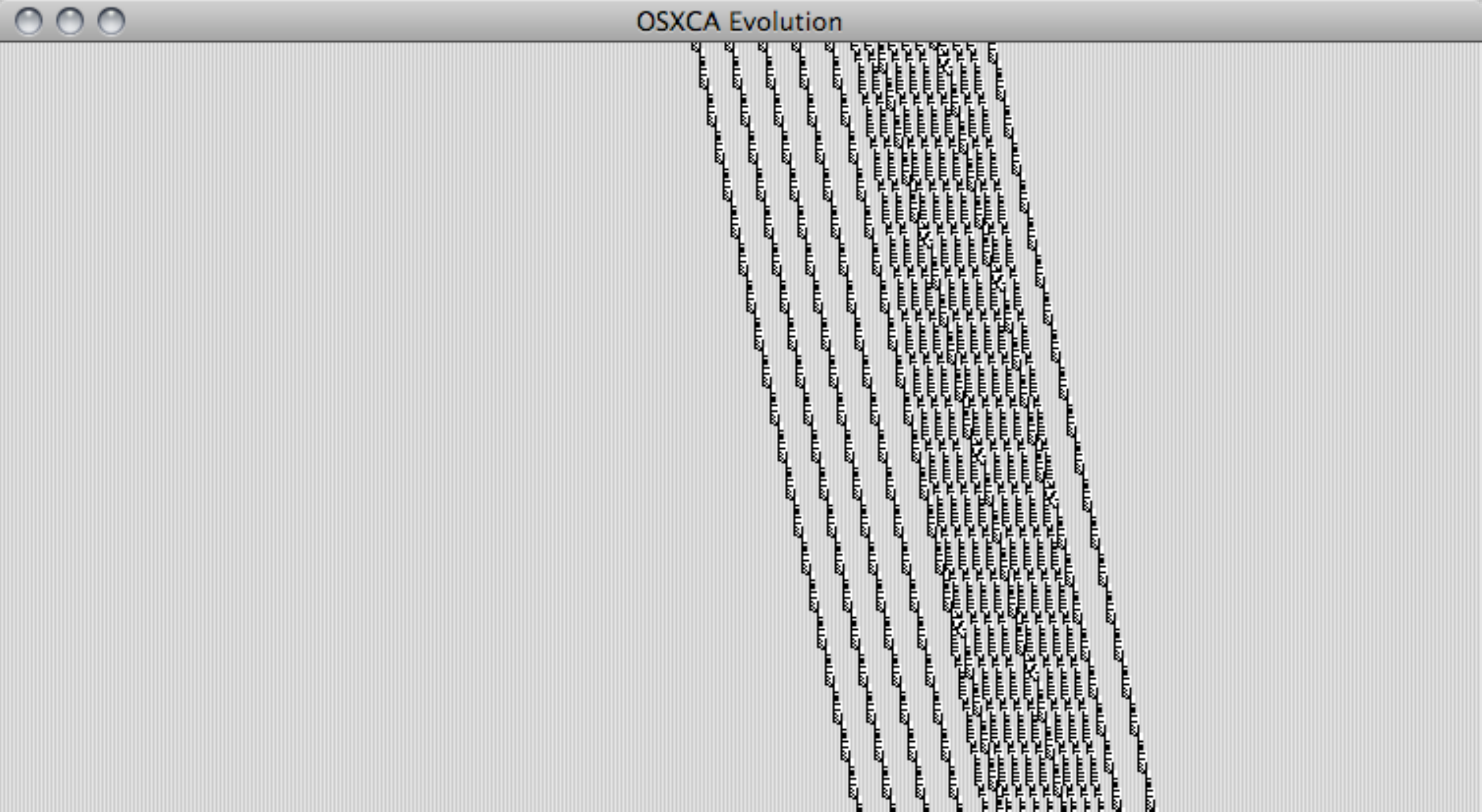}}} \hspace{0.7cm}
\subfigure[]{\scalebox{0.35}{\includegraphics{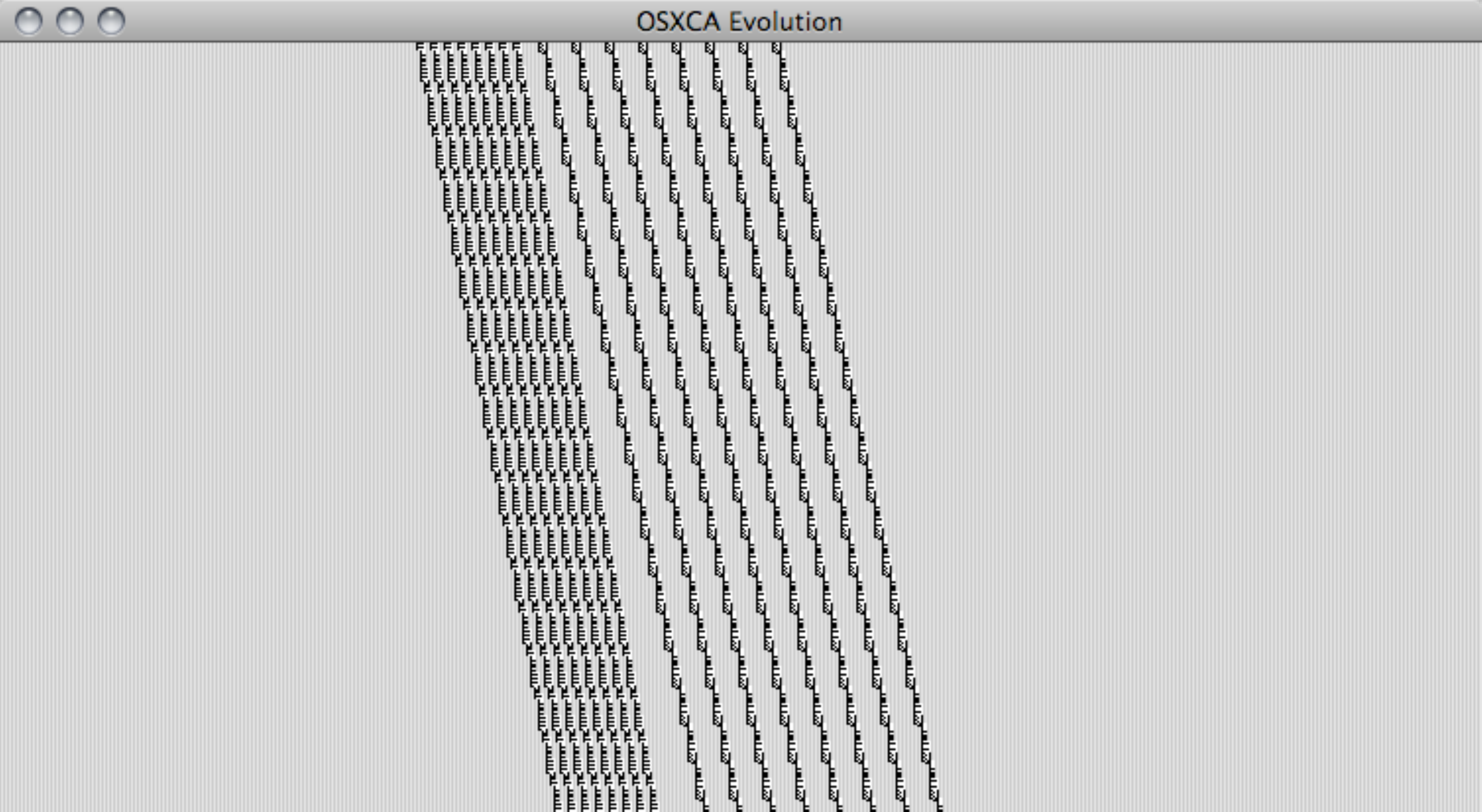}}}
\caption{A simple substitution system processing the word $A^{8}B^{8}$ to $B^{8}A^{8}$ with $\phi_{R30maj:8}$. The final required state is reached at the 5,000th generation by synchronization of multiple soliton reactions.}
\label{substitution}
\end{figure}

Let us consider a ECA with memory, rule $\phi_{R30maj:8}$~\cite{kn:MAA10}. We want to implement a simple substitution function {\sf addToHead} working on two strings $w_1=A_1,\ldots,A_n$ and $w_2=B_1,\ldots,B_m$, where $n,m \geq 1$. For example, if $w_1=AAA$, $w_2=BBB$ and $w_3=w_1w_2$ then {\sf addToHead($|w_2|$)} means produce $w_3=w_2w_1$. To implement such a function in $\phi_{R30maj:8}$ every quantum of data is represented by a particle. Particles $g_1$ and $g_2$ are coded in order to reproduce a soliton reaction. The codification is not sophisticated. However, a systematic analysis of reactions is required. A periodic gap and one fixed phase between particles is sufficient to reproduce {\sf addToHead} function for any string $A^nB^m$.

Figure~\ref{substitution} shows the evolution of $\phi_{R30maj:8}$ starting from an initial condition coded by particles representing the string $AAAAAAAABBBBBBBB$. Using function {\sf addToHead} we produce the final string $BBBBBBBBAAAAAAAA$ after 5,000 generations. The first snapshot (Fig.~\ref{substitution}a) shows the initial configuration and the first 391 generations, the middle snapshot (Fig.~\ref{substitution}b) demonstrates how string $w_1$ acts on string $w_2$ while preserving the information (soliton or identity reaction), and the third snapshot (Fig.~\ref{substitution}c) shows the final global configuration (i.e. string $w_2w_1$ processed in parallel by $\phi_{R30maj:8}$).

\begin{figure}[th]
\centerline{\includegraphics[width=0.93in]{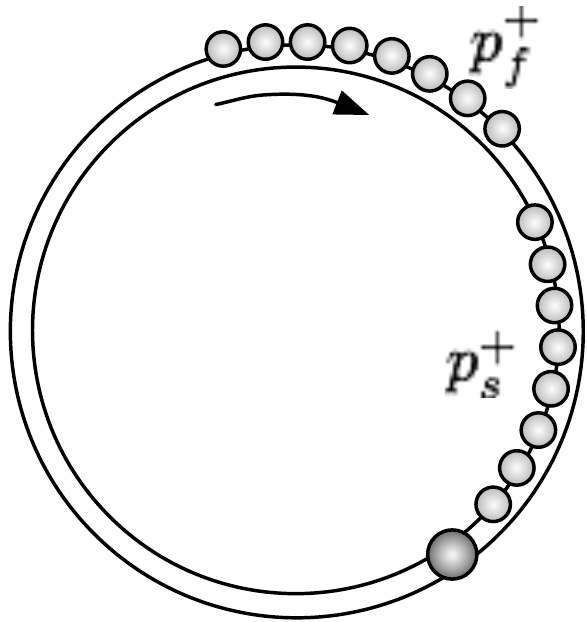}}
\caption{Beam routing performing identity reactions in $\phi_{R30maj:8}$. A cycle realizes its operation, two cycles reinitialize the beam state and the operations can then be repeated.}
\label{beamRoutingR30maj8}
\end{figure}

Therefore a beam routing to represent such an operation in $\phi_{R30maj:8}$ requires two particles (as positrons but one slow and other fast respectively) with the same orientation and one collision contact point on the beam routing. Thus Fig.~\ref{beamRoutingR30maj8} illustrates how a beam routing is designed to produce such periodic collisions.

\begin{figure}[th]
\centerline{\includegraphics[width=3.3in]{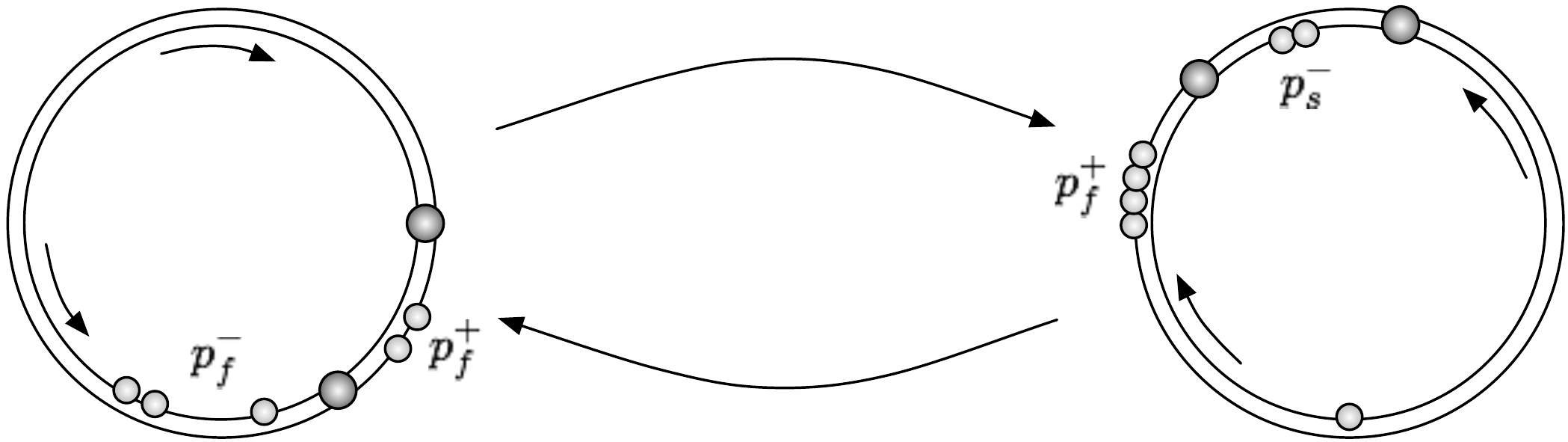}}
\caption{Transition between two beam routing synchronizing multiple reactions. When the first set of collisions are done a new beam routing is defined with other particles, so that when the second set of collisions is done then one returns to the initial condition of the first beam, constructing a meta-glider or mesh in $\varphi_{R110}$.}
\label{beamRoutingTransition}
\end{figure}

\begin{figure}[th]
\centerline{\includegraphics[width=4.5in]{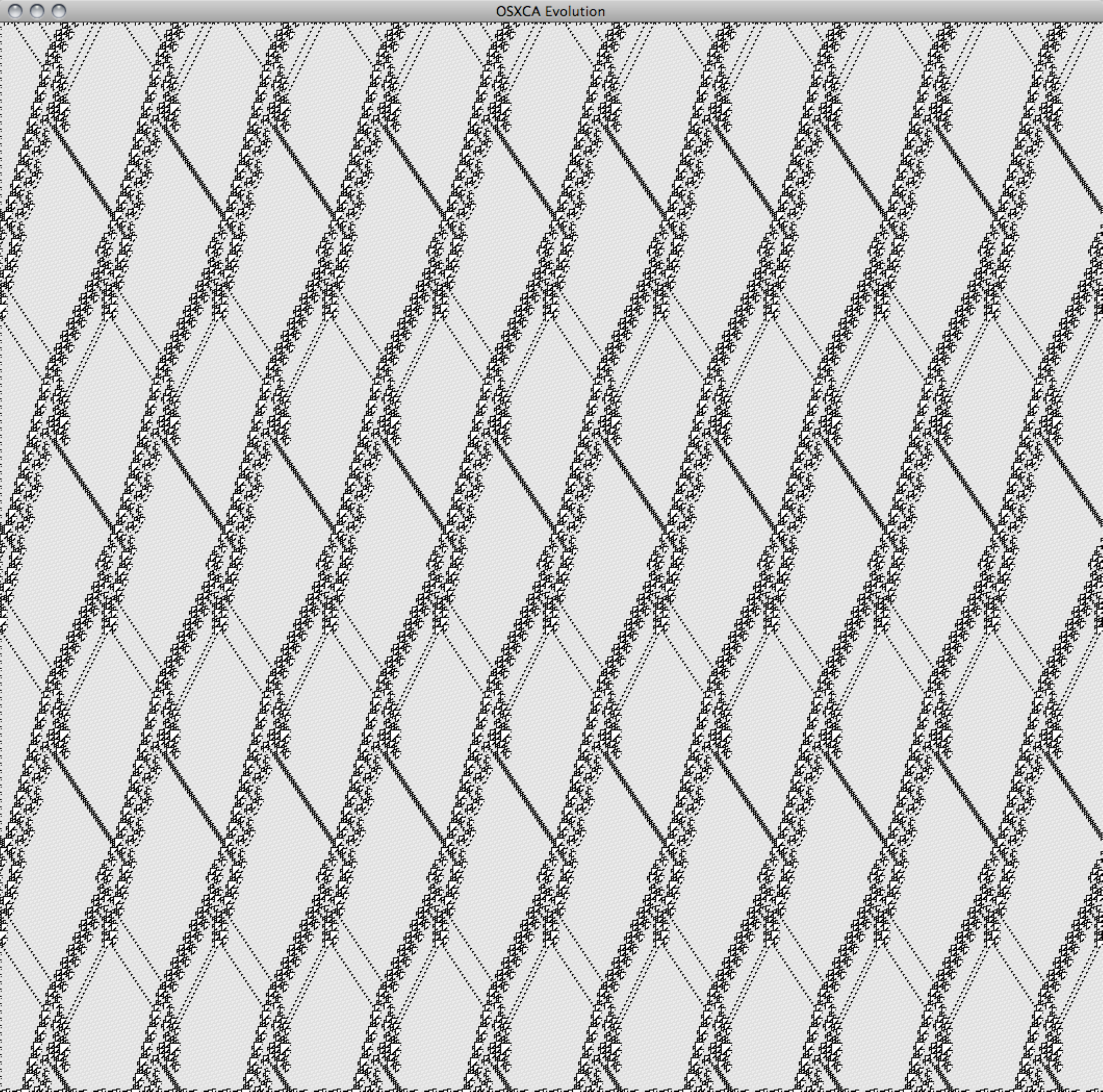}}
\caption{Synchronization of multiple collisions in $\varphi_{R110}$ on a ring of 1,060 cells in 1,027 generations, starting with 50 particles from its initial condition.}
\label{metaR110}
\end{figure}

In this way, we can design more complex constructions synchronizing multiple collisions with a diversity of speeds and phases on different particles. Figure~\ref{beamRoutingTransition} displays a more sophisticated beam routing design, connecting two of beams and then creating a new beam routing diagram where edges represent a change of particles and collisions contact point on ECA $\varphi_{R110}$. In such a transition, a number of new particles emerge and collide to return to the first beam, thus oscillating between two beam routing forever. To better understand this double beam routing dynamic, consider Fig.~\ref{metaR110} where we see multiple collisions between particles (first beam routing):

$$
p_{A}^+, p_{A}^+ \leftrightarrow p_{\bar{B}}^-, p_{B}^-, p_{B}^-
$$

\noindent changing to the set of particles (second beam routing):

$$
p_{A^4}^+ \leftrightarrow p_{E}^+, p_{\bar{E}}^+
$$

\noindent defining two beam routing connected by a transition of collisions as:

$$
(p_{A}^+, p_{A}^+ \leftrightarrow p_{\bar{B}}^-, p_{B}^-, p_{B}^-) \rightarrow (p_{A^4}^+ \leftrightarrow p_{E}^+, p_{\bar{E}}^+) \mbox{, and }$$
$$
(p_{A^4}^+ \leftrightarrow p_{E}^+, p_{\bar{E}}^+) \rightarrow (p_{A}^+, p_{A}^+ \leftrightarrow p_{\bar{B}}^-, p_{B}^-, p_{B}^-).
$$

So we see that a beam routing representation not only helps in designing collisions but also to implement computation. 

We now explain the function of a cyclic tag system (CTS) in $\varphi_{R110}$ in beam routing terms and discuss a number of theoretical implications. 

It is well known that ECA $\varphi_{R110}$ is universal~\cite{kn:Cook04, kn:Wolf02}. However, a number of details about its construction, e.g. a cyclic machine, are uncertain~\cite{kn:MMS11}.\footnote{A complete description of CTS working in $\varphi_{R110}$ is available at  \url{http://uncomp.uwe.ac.uk/genaro/rule110/ctsRule110.html}} Simplified implementations of universal computation in CTS are provided in~\cite{kn:Cook08, kn:NW06}.

\begin{table}[th]
\centering
\small
\begin{tabular}{|c|c|}
\hline
particle & velocity \\
\hline \hline
$A$ & 2/3 $\approx$ 0.666666  \\
\hline
$B$ & -2/4 = -0.5 \\
\hline
$\bar{B}^{n}$ & -6/12 = -0.5 \\
\hline
$\hat{B}^{n}$ & -6/12 = -0.5 \\
\hline
$C_{1}$ & 0/7 = 0 \\
\hline
$C_{2}$ & 0/7 = 0 \\
\hline
$C_{3}$ & 0/7 = 0 \\
\hline
$D_{1}$ & 2/10 = 0.2 \\
\hline
$D_{2}$ & 2/10 = 0.2 \\
\hline
$E^{n}$ & -4/15 $\approx$ -0.266666 \\
\hline
$\bar{E}$ & -8/30 $\approx$ -0.266666 \\
\hline
$F$ & -4/36  $\approx$ -0.111111 \\
\hline
$G^{n}$ & -14/42 $\approx$ -0.333333 \\
\hline
$H$ & -18/92 $\approx$ -0.195652 \\
\hline
gun & -20/77 $\approx$ -0.259740 \\
\hline
\end{tabular}
\caption{Properties of particles in $\varphi_{R110}$.}
\label{margenesgliders}
\end{table}

ECA $\varphi_{R110}$ has a unique complexity due to great number of particles that emerge in the automaton evolution. Table~\ref{margenesgliders} presents a summary of the basic particle properties in $\varphi_{R110}$, following Cook's nomenclature. We can appreciate how diverse they are. There are even particles that can expand size forever which increases the number of collisions that we can produce on this ECA.

Initially, a CTS works as a traditional tag system \cite{kn:Mins67}, i.e., as  a substitution system reading the first symbol on the tape, deleting and putting new symbols. CTS are new machines proposed by Cook~\cite{kn:Cook04} as a tool to implement computations in $\varphi_{R110}$. CTS are a variant of tag systems: they have the same action of reading a tape in the front and adding characters at the end, nevertheless there are some new features as follows:

\begin{enumerate}
\item A CTS needs at least two letters in its alphabet ($\mu > 1$).
\item Only the first character is deleted ($\nu = 1$) and its respective sequence is added.
\item If the machine reads a character zero then the production rule is always null ($0 \rightarrow \epsilon$, where $\epsilon$ represents the empty word).
\item There are $k$ sequences from $\mu^*$ which are periodically accessed to specify the current production rule when a nonzero character is taken by the system. Therefore the period of each cycle is determinate by $k$.
\end{enumerate}

This way a cycle determines a partial computation over the tape. No particular halt conditions are specified in the original paper~\cite{kn:Cook04}, possibly because the halting is a direct consequence of tag systems. Let us see some samples of a CTS working with $\mu=2$, $k=3$ and the production rules: $1 \rightarrow 11$, $1 \rightarrow 10$ and $1 \rightarrow \epsilon$.

\begin{figure}[th]
\centerline{\includegraphics[width=4.5in]{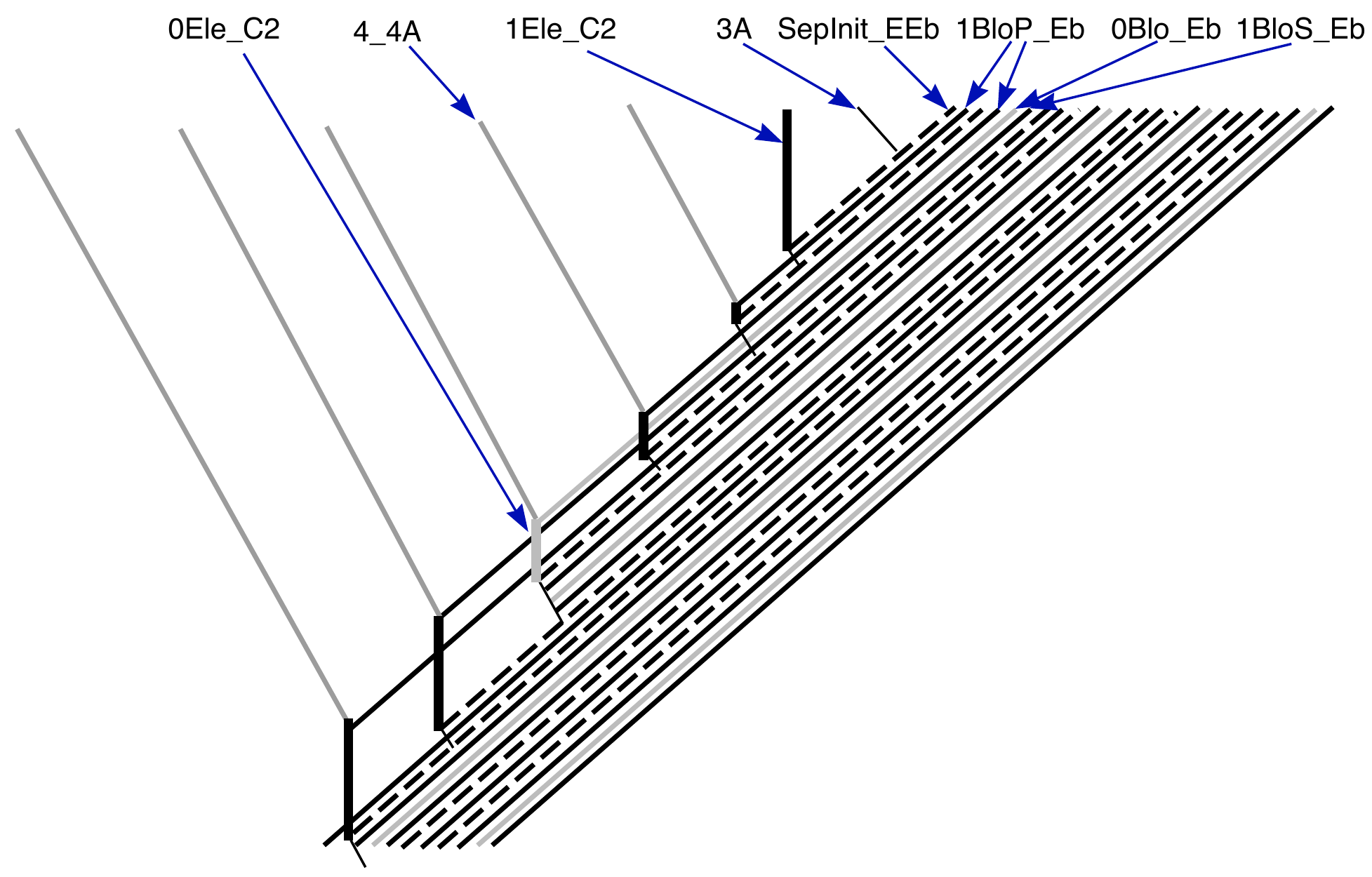}}
\caption{A diagram of a CTS working in $\varphi_{R110}$.}
\label{CTS-diagram}
\end{figure}

Our intention is to represent CTS function in $\varphi_{R110}$ based beam routing of symbols. We employ transition beam routings to explain critical changes that occur due to collisions and sets of particles that are present on each beam routing at every instant description of the machine.

We start with a description of main stages of encoding packages of particles in their initial condition. This will be an equivalent of beam routing diagram. We will first show how particles and their collisions emulate a CTS in $\varphi_{R110}$. We use packages of particles to represent data and operators. We read, transform and delete data in the tape using reactions between the particles. The approach is a delicate one and requires a laborious task of setting initial configurations for the particles in the beams. A diagram of such representation is shown in Fig.~\ref{CTS-diagram}, particular features of the diagram are explain below.

A construction of the CTS in $\varphi_{R110}$ can be partitioned, essentially, in three parts:

\begin{itemize}
\item First, is the left periodic part, controlled by packages of 4\_$A^{4}$ particles. This part is static and controls the production of $0$'s and $1$'s. 
\item The second part is the center, determining the initial value in the tape.
\item The third one is the right cyclic part, which has the data to process, adding a leader component which specifies data added to or erased from the tape in the evolution space.
\end{itemize}

\begin{figure}
\centering
\subfigure[]{\scalebox{0.62}{\includegraphics{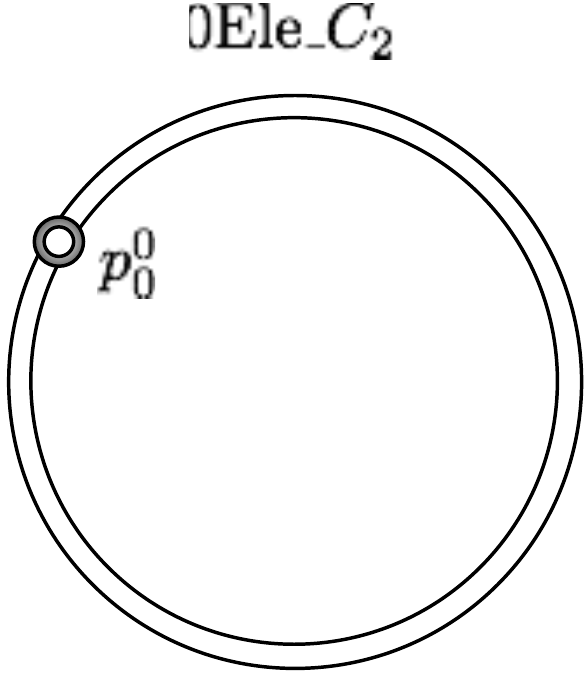}}} \hspace{0.2cm}
\subfigure[]{\scalebox{0.62}{\includegraphics{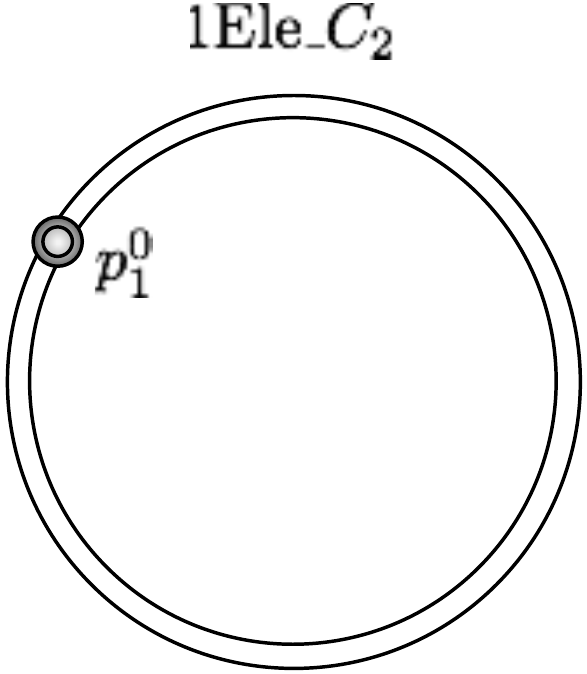}}} \hspace{0.2cm}
\subfigure[]{\scalebox{0.62}{\includegraphics{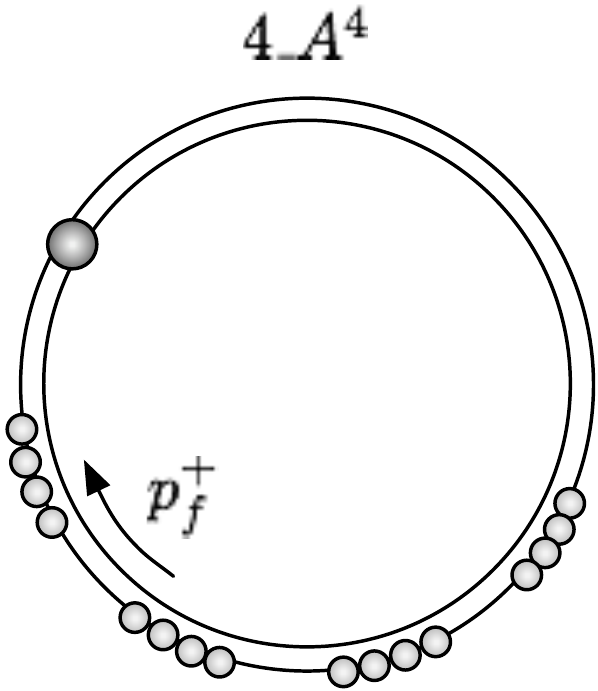}}} \hspace{0.5cm}
\subfigure[]{\scalebox{0.62}{\includegraphics{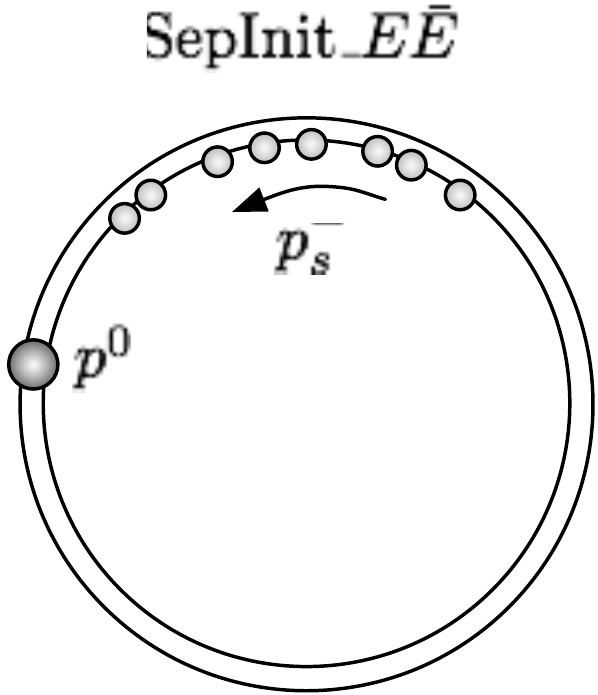}}} \hspace{0.2cm}
\subfigure[]{\scalebox{0.62}{\includegraphics{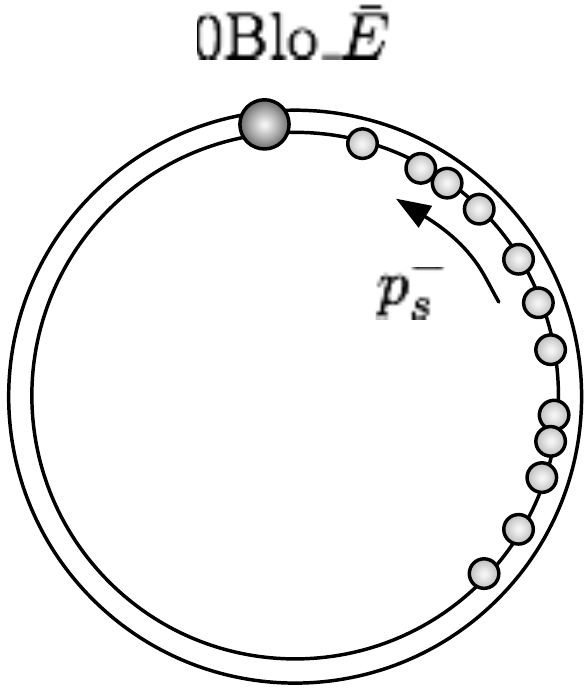}}} \hspace{0.2cm}
\subfigure[]{\scalebox{0.62}{\includegraphics{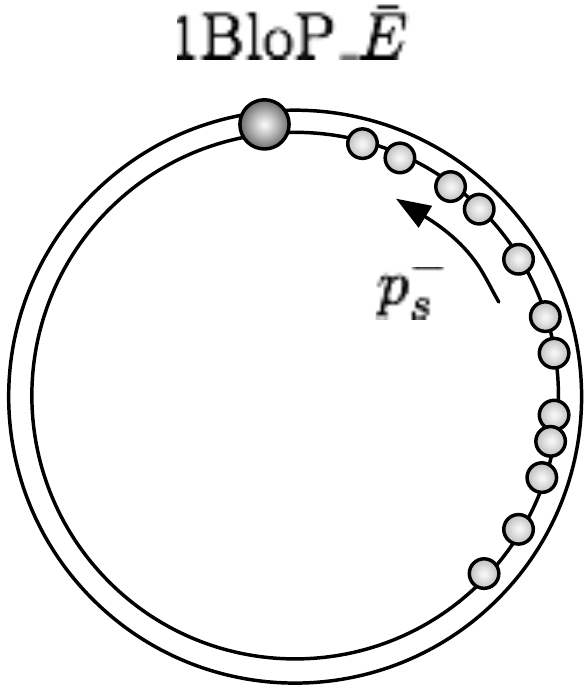}}} \hspace{0.5cm}
\subfigure[]{\scalebox{0.62}{\includegraphics{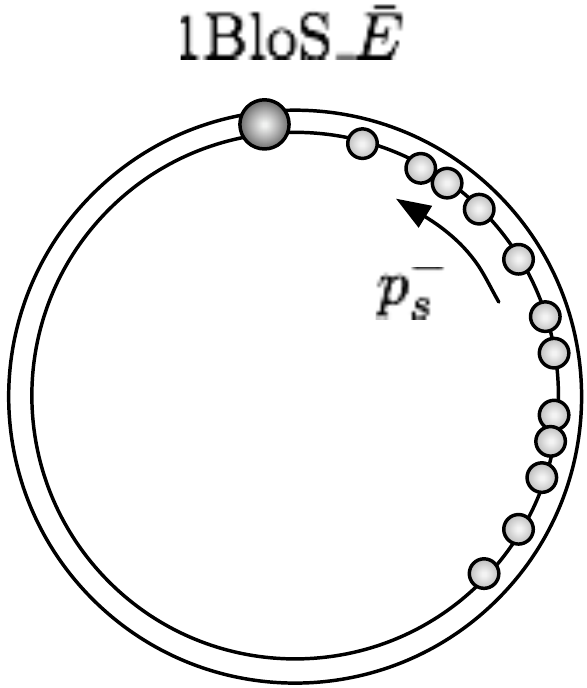}}} \hspace{0.2cm}
\subfigure[]{\scalebox{0.62}{\includegraphics{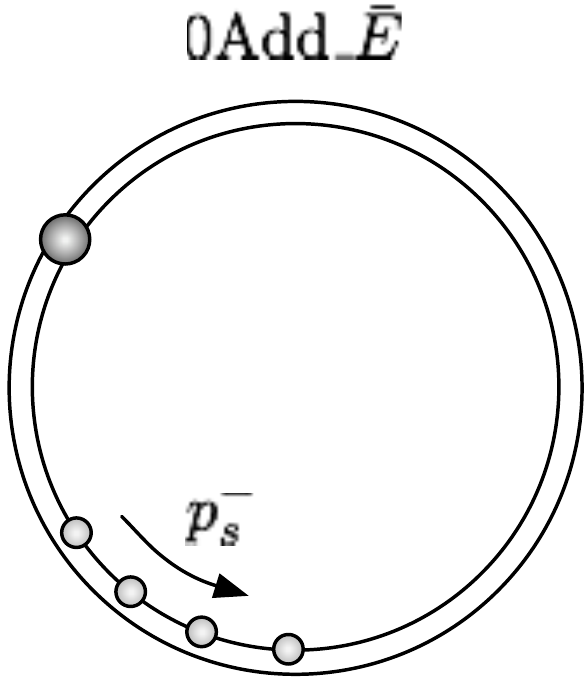}}} \hspace{0.2cm}
\subfigure[]{\scalebox{0.62}{\includegraphics{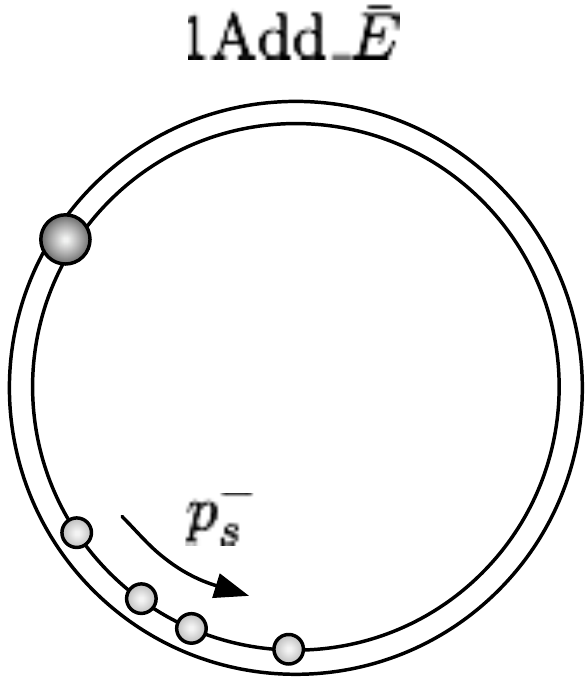}}} 
\caption{Beam routing codification representing package of particles which reproduces a CTS in $\varphi_{R110}$.}
\label{beamRoutingR110}
\end{figure}

In the left part, the four packages of $A^{4}$ (Fig.~\ref{beamRoutingR110}c) particles must be carefully explained because although they are static, their phases change periodically. The important point is to implement these components in defining both distances and phases, because a distinct phase or a distance induces an undesirable reaction; in fact all components obey this restriction because every glider of each component must be correctly aligned. 

The central part is constituted by an initial 1 on the tape represented by a package of four $C_{2}$ particles. This way, an element 1Ele$\_C_{2}$ (Fig.~\ref{beamRoutingR110}b) represents a 1 and the element 0Ele$\_C_{2}$ (Fig.~\ref{beamRoutingR110}a) represents a 0 in the tape.

The element 0Blo$\_\bar{E}$ is formed by 12$\bar{E}$ particles. For the 1Blo$\_\bar{E}$ element, Cook establishes the existence of two components to represent 1's: 1BloP$\_\bar{E}$ (Fig.~\ref{beamRoutingR110}f) named {\it primary} and 1BloS$\_\bar{E}$ (Fig.~\ref{beamRoutingR110}g) named {\it standard}. In essence both blocks produce the same element 1Add$\_\bar{E}$, although coming from different intervals. The reason to use both blocks is that $\varphi_{R110}$ is not symmetric; thus, if we use only one element then although we would obtain a good production in the first collision, generating an element 1Add$\_\bar{E}$, when the second package of 4\_$A^{4}$ particles finds the second block, the system is completely destroyed.

A leader element SepInit$\_E\bar{E}$ (Fig.~\ref{beamRoutingR110}d) is important in order to separate packages of data and determine their incorporation into of the tape.

The elements 1Add$\_\bar{E}$ (Fig.~\ref{beamRoutingR110}i) and 0Add$\_\bar{E}$ (Fig.~\ref{beamRoutingR110}h) are produced by two previous different packages of data. An element 1Add$\_\bar{E}$ must be generated by a block 1BloP$\_\bar{E}$ or by 1BloS$\_\bar{E}$. This way, both elements can produce the same element. While an element 0Add$\_\bar{E}$ is generated by a block 0Blo$\_\bar{E}$ (Fig.~\ref{beamRoutingR110}e).

Finally, in terms of periodic phases, this CTS can be written as follow:

\begin{itemize}
\item[{\bf left:}] \{$649e$-4$\_A^{4}$(F$\_{i}$)\}*, for $1 \leq i \leq 3$ in sequential order
\item[{\bf center:}] $246e$-1Ele$\_$C2(A,f$_{1}$\_1)-$e$-$A^{3}$(f$_{1}$\_1)
\item[{\bf right:}] \{SepInit$\_E\bar{E}$(\#,f$_{i}$\_1)-1BloP$\_\bar{E}$(\#,f$_{i}$\_1)-SepInit$\_E\bar{E}$(\#,f$_{i}$\_1)-1BloP$\_\bar{E}$(\#,f$_{i}$\_1)-0Blo$\_\bar{E}$(\#,f$_{i}$\_1)-1BloS$\_\bar{E}$(\#,f$_{i}$\_1)\}* (where $1 \leq i \leq 4$ and \# represents a particular phase).
\end{itemize}

For a complete and full description of such reproduction by phases f$_{i}\_1$, please see \cite{kn:MMS11}.

To get a CTS emulation in $\varphi_{R110}$ by beam routings, we will use connections between beam routing, as we have proposed in Fig~\ref{beamRoutingTransition}.

\begin{figure}
\centerline{\includegraphics[width=3.6in]{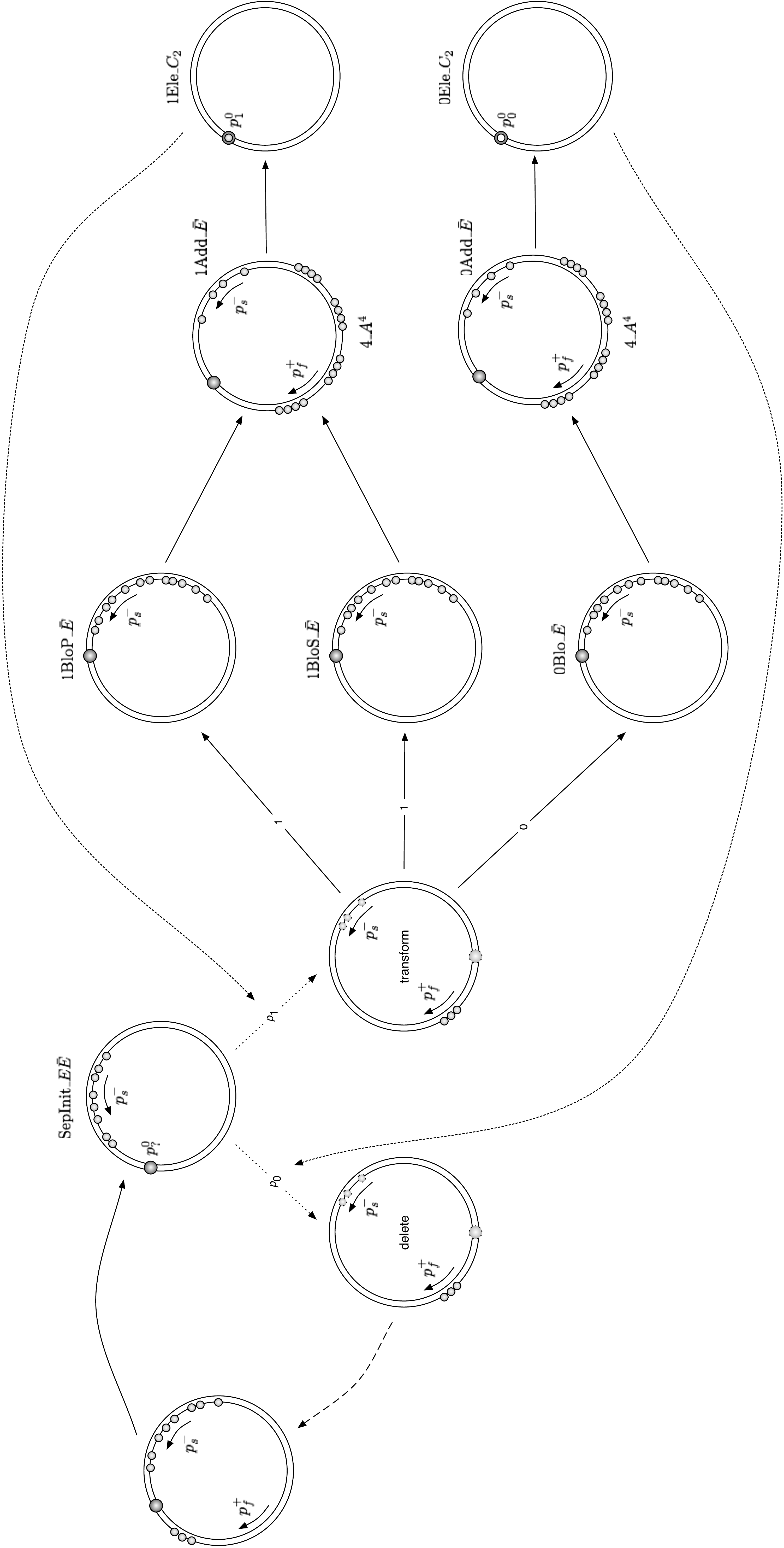}}
\caption{Beam routing machine transitions to simulate CTS in $\varphi_{R110}$.}
\label{CTSbeamrouting}
\end{figure}

Transitions between beam routings connect each set of collisions to enter to the next beam routing diagram.

We need some fine details to get a CTS operation by beam routings. Figure~\ref{CTSbeamrouting} shows the general diagram to reproduce a CTS by particle collisions in $\varphi_{R110}$ with beam routing transitions. While Fig.~\ref{beamRoutingR110} describes each component to code particles in $\varphi_{R110}$, Fig.~\ref{CTSbeamrouting} display how to connect such components and control the transition of collisions.

Some notes are necessary to better understand this schematic diagram. For the components 1Ele$\_C_2$ and 0Ele$\_C_2$ they are compressed only for one dark circle (that represents the point of collisions). Both elements are constituted for four $C_2$ particles in different distances, although they have a static position $p^0$ that can be confined to this dark circle, as $p_1^0$ and $p_0^0$ respectively.

When a leader component (SepInit$\_E\bar{E}$) is transformed, given previous binary value on the tape, it collides with $p_?^0$ component, i.e. a $p_1^0$ or $p_0^0$ element. If $p_?^0$ is 0, then a cascade of collisions start to {\it delete} all components that come with three particles successively and change this reaction to reach a new leader component and return to the beginning. But, if $p_?^0$ is 1 then a cascade of {\it transformations} dominated by additional particles $p^0$ starts, in order to reach the next leader component. Here, we have more variants because pre-transformed package of particles is encoded to binary values that is then included on the machine tape. If a block of particles is 1BloP$\_\bar{E}$ or 1BloS$\_\bar{E}$ this means that such a component will be transformed to one 1Add$\_\bar{E}$ element. But, if a block of particles is 0Blo$\_\bar{E}$, then such a component will be transformed to 0Add$\_\bar{E}$ element. At this stage, when both components are prepared then a binary value is introduced on the tape, a 1Add$\_\bar{E}$ element stores a 1 (1Ele$\_C_2$), and a 0Add$\_\bar{E}$ element stores a 0 (0Ele$\_C_2$), which eventually is deleted for the next leader component and starts a new cycle in the CTS machine.

The relevant point here is how to control and code particles by beam routings. That will offer a better chance to perform a computation with this architecture before implementing the complicated and laborious task of setting up initial condition. Of course, such an idea can be carried out in any CA handling signals, waves, particles, gliders or mobile self-localizations \cite{kn:Ada01, kn:Ada02, kn:Ada02a, kn:Akin99, kn:CD98, kn:JSS01, kn:Koh89, kn:Kurka03, kn:LPM07, kn:MAM06, kn:Marg84, kn:Marg98, kn:Marg03, kn:MMI99, kn:MNG04, kn:Mor08, kn:Piva07, kn:SCJ09, kn:Hey98}.

The package particles codification for a CTS in \cite{kn:Wolf02, kn:MMS11} is represented for the expression $(s1s101)^+$ while that in \cite{kn:Cook04} is represented for the expression $(s111s0)^+$. Finally in all cases, this CTS beam routing may simulate such operations.

\section{Summary}
\label{summary}

We advanced a concept of symbol super collider~\cite{kn:Toff02} and implemented it in one-dimensional cellular automata. Two types of automaton rings were considered --- elementary cellular automata (binary cell state, ternary neighbourhoods) and elementary cellular automata with memory. We demonstrated that, for certain rules of cell-state transition functions, the automata exhibit a wide range of particles (mobile self-localizations or gliders) in their evolution. High degree of morphological diversity and richness of collision dynamics of the particles in memory-enriched cellular automata make them particularly promising objects for constructing computational schemes. 

In present paper we advanced over twenty years old studies in 
identifying and classifying gliders and localisations in CA. There is a
number of approaches varying in their degrees of efficiency and discovery potentials.
Thus, Wuensche successfully classified CA evolution rules with his Z parameter,
and provided power structures of a substantial number of glider-generating rules~\cite{kn:Wue99}. 
Other studies focus on specification of sets of strings as patterns and gliders. For example, 
Redeker developed a compact algebraic representation of 1D CA evolution, known as {\it
flexible time}. Suing his approach one can construct specific periodic structures~\cite{kn:Red10}. 
Evolution programming techniques are also proved to be very successful, e.g. in situations 
when a genetic algorithm undertakes a search for gliders and patterns on fixed small arrays of 
bits~\cite{kn:SBC07, kn:WO08}. As well, Freire {\it et. al} developed an approach to select specific
sets of vectors, initial configurations, that determine development of gliders and localised 
patterns~\cite{kn:FBO07, kn:FBO10}. Finally, the de Bruijn diagrams is a time-tested and reliable tool 
to classify sets of periodic structures or gliders using exhaustive approach~\cite{kn:Mc90, kn:Mc09}.
Said that in present paper we did not aim to deal with identification of mobile patterns but 
rather consider glider sets in collision-based computing perspective. Concretely, we only aimed to 
develop the idea of cyclotrons as an abstract model of super-colliders~\cite{kn:Toff02}.

We proved experimentally that the dynamics of these particles support all basic types of ballistic collisions and consequently that the cellular automata cyclotrons can be used to implement certain schemes and operations of collision-based computing. We designed schemes with several beams of particles and provided a means of beam routing and programming interactions between particles. To demonstrate the high degree of computational universality of cellular automaton 'cyclotrons' we took into account that each particle is essentially a finite-size binary string travelling along a one-dimensional lattice `cyclotron.'  When a string with lower velocity overtaken by a string with higher velocity the content of one or both strings can be modified. Thus, for carefully selected initial configuration of particles and types of particles on a ring, the cellular automaton cyclotron imitates a cyclic tag systems, and thus is computationally universal. 

What are advantages of the proposed approach?  In cellular-automaton models particles can run along the rings indefinitely. With suitable beam routing schemes some particles (the results of computation or the products of reactions) can be removed from the system and new particles can be added. The cellular-automaton super collider is a universal computing devices based on interactions between particles due to the particles' different speeds. This give our constructs an enormous advantage when compared to more ``traditional'' constructs, where particles collide only due to different orientations of velocity vectors. Also most existing unconventional designs of universal computing devices have certain issues of boundary conditions, e.g. infinite versus finite, the cellular-automaton super-colliders by using periodic boundary conditions removes the effect of any boundary being equivalent to an infinite periodic system.

The theoretical models of cellular-automaton super colliders open new perspectives in laboratory implementations of collision based computing devices. The opportunities are virtually endless. Hundreds of chemical, physical and biological systems exhibit travelling localizations in their dynamics. Examples include co-aligned dipole groups in tubulin microtubules (e.g. \cite{rasmussen_1990,mershin_2004}, kinks, breathers and solitons in molecular chains and polymers (e.g. ~\cite{davydov_1990},  phasons in quasi-crystals~\cite{lipp_2010}, kinks in ferromagnets~\cite{coldea_2010}, dissipative solitons in gas-discharge systems~\cite{astrov_2009}, localizations of electron density in monolayers of aromatic molecules~\cite{bandyo_2010}, wave-fragments in excitable chemical media~\cite{kn:ACA05}. We envisage that molecular chains and polymers would be the best candidates for experimental laboratory realisation of symbol super colliders. This will be a subject of further studies.

\section*{Acknowledgement}
Genaro J. Mart{\'i}nez thanks to support given by DGAPA-UNAM and EPSRC grant EP/F054343/1.



\begin{thebibliography}{99}

\bibitem{kn:Ada01} Adamatzky, Andrew, {\em Computing in Nonlinear Media and Automata Collectives}, Institute of Physics Publishing, Bristol and Philadelphia 2001.

\bibitem{kn:Ada02} Adamatzky, Andrew (Ed.) {\em Collision-Based Computing}, Springer 2002.

\bibitem{kn:Ada02a} Adamatzky, Andrew, ``New media for collision-based computing,'' In \cite{kn:Ada02}, 411--442, 2002.

\bibitem{kn:ACA05} Adamatzky, Andrew, Costello, Ben L., and Asai, Tetsuya, {\em Reaction-Diffusion Computers}, Elsevier 2005.

\bibitem{kn:AM03} Alonso-Sanz, R. and Martin, M., ``Elementary CA with memory,'' {\em Complex Systems} {\bf 14} 99--126, 2003.

\bibitem{kn:Alo06} Alonso-Sanz, R., ``Elementary rules with elementary memory rules: the case of linear rules,'' {\em Journal of Cellular Automata} {\bf 1} 71--87, 2006.

\bibitem{kn:Alo09} Alonso-Sanz, Ramon, {\em Cellular Automata with Memory}, Old City Publishing 2009.

\bibitem{astrov_2009} Astrov~Y., ``Gas-discharge planar semiconductor structures as devices for un-conventional computing,'' {\it Int. J. Unconventional Computing}, in press.

\bibitem{kn:Akin99} Awazu, Akinori, ``Cellular automaton rule 184++C. A simple model for the complex dynamics of various particles flow,'' {\em Physics Letters A} {\bf 261(5-6)} 309--315, 1999.

\bibitem{bandyo_2010} Bandyopadhyay~A., Pati~R., Sahu~S., Peper~F., and Fujita~D., ``Massively parallel computing on an organic molecular layer,'' {\it Nature Physics} {\bf 6} 369--375, 2010. 

\bibitem{kn:BNR91} Boccara, Nino, Nasser, J., and Roger, M. ``Particle like structures and their interactions in spatio-temporal patterns generated by one-dimensional deterministic cellular automaton rules,'' {\em Physical Review A} {\bf 44(2)} 866--875, 1991.

\bibitem{kn:CD98} Chopard, Bastien and Droz, Michel, {\em Cellular Automata Modeling of Physical Systems}, Collection Al\'ea Saclay, Cambridge University Press 1998.

\bibitem{kn:Cook04} Cook, Matthew, ``Universality in Elementary Cellular Automata,'' {\em Complex Systems} {\bf 15(1)} 1--40, 2004.

\bibitem{kn:Cook08} Cook, Matthew, ``A Concrete View of Rule 110 Computation,'' In {\em The Complexity of Simple Programs} (T. Neary, D. Woods, A. K. Seda, and N. Murphy (Eds.)), 31--55, 2008.

\bibitem{coldea_2010} R. Coldea, D. A. Tennant, E. M. Wheeler, E. Wawrzynska, D. Prabhakaran, M. Telling, K. Habicht, P. Smeibidl, and K. Kiefer, ``Quantum criticality in an Ising chain: Experimental evidence for emergent E8 symmetry,'' {\em Science} {\bf 327} 177--180, 2010.

\bibitem{davydov_1990} Davydov~A.~S., {\em Solitons in Molecular Systems}, Springer 1990.

\bibitem{kn:FT01} Fredkin, Edward and Toffoli, Tommaso, ``Design Principles for Achieving High-Performance Submicron Digital Technologies,'' In \cite{kn:Ada02}, 27--46, 2001.

\bibitem{kn:FBO07} Freire, Joana G., Brison, Owen J., and Gallas,
Jason A.C. (2010) ``Spatial updating, spatial transients, and
regularities of a complex automaton with nonperiodic architecture,''
{\em Chaos} {\bf 17} 026113.

\bibitem{kn:FBO10} Freire, Joana G., Brison, Owen J., and Gallas,
Jason A.C. (2010) ``Complete sets of initial vectors for pattern growth
with elementary cellular automata,'' {\em Computer Physics
Communications} {\bf 181} 750--755.

\bibitem{fuerstman_2003} Michael J. Fuerstman, Pascal Deschatelets, Ravi Kane, Alexander Schwartz, Paul J. A. Kenis, John M. Deutch, and George M. Whitesides, ``Solving mazes using microfluidic networks,'' {\em Langmuir} {\bf 19} 4714--4722, 2003.

\bibitem{gorecki_2003} G\'{o}recki J., Yoshikawa K., and Igarashi Y., 
``On chemical reactors that can count,'' {\em J. Phys. Chem. A} {\bf 107}  1664--1669, 2003.

\bibitem{kn:HC97} Hanson, James E. and Crutchfield, James P., ``Computacional Mechanics of Cellular Automata: An Example,'' {\em Physics D} {\bf 103} 169--189, 1997.

\bibitem{kn:Hey98} Hey, Anthony J. G., {\em Feynman and computation: exploring the limits of computers}, Perseus Books 1998.

\bibitem{kn:JSS01} Jakubowski, Mariusz H., Steiglitz, Kenneth, and Squier, Richard, ``Computing with Solitons: A Review and Prospectus,'' {\em Multiple-Valued Logic} {\bf 6(5-6)} 439--462, 2001.

\bibitem{katz_2010} Katz~E. and Privman~V., ``Enzyme-based logic systems for information processing,'' {\em Chem. Soc. Rev.} {\bf 39} 1835--1857, 2010. \url{arXiv:0910.0270}

\bibitem{kn:Koh89} Kohyama, Tamotsu, ``Cellular automata with particle conservation,'' {\em Progress of Theoretical Physics} {\bf 81(1)} 47--59, 1989.

\bibitem{kn:Kurka03} K\r{u}rka, Petr, ``Cellular automata with vanishing particles,'' {\em Fundamenta Informaticae} {\bf 58(3-4)} 203--221, 2003.

\bibitem{lipp_2010} Lipp~G., Engel~M., Sonntag~S., and Trebin~H. R., Phason, ``Dynamics in One-Dimensional Lattices,'' \url{arXiv:1002.1918v1}, 2010.

\bibitem{kn:LPM07} Lizier, Joseph T., Prokopenko, Mikhail, and Zomaya, Albert Y., ``Information transfer by particles in cellular automata,'' {\em Lecture Notes in Computer Science} {\bf 4828} 49--60, 2007.

\bibitem{kn:MAA10} Mart\'{\i}nez, Genaro J., Adamatzky, Andrew, Alonso-Sanz, Ramon, and Seck-Tuoh-Mora, J. C., ``Complex dynamic emerging in Rule 30 with majority memory,'' {\em Complex Systems} {\bf 18(3)} 345--365, 2010.

\bibitem{kn:MAS10} Mart\'{\i}nez, Genaro J., Adamatzky, Andrew, Seck-Tuoh-Mora, J. C., and Alonso-Sanz, Ramon, ``How to make dull cellular automata complex by adding memory: Rule 126 case study,'' {\em Complexity} {\bf 15(6)} 34--49, 2010.

\bibitem{kn:MAM06} Mart\'{\i}nez, Genaro J., Adamatzky, Andrew, and McIntosh, Harold V., ``Phenomenology of glider collisions in cellular automaton Rule 54 and associated logical gates,'' {\em Chaos, Solitons and Fractals} {\bf 28} 100--111, 2006.

\bibitem{kn:MAM08} Mart{\'i}nez, Genaro J., Adamatzky, Andrew, and McIntosh, Harold V., ``On the representation of gliders in Rule 54 by de Bruijn and cycle diagrams,'' {\em Lecture Notes in Computer Science} {\bf 5191} 83--91, 2008.

\bibitem{kn:Marg84} Margolus, Norman, ``Physics-like models of computation,'' {\em Physica D} {\bf 10(1-2)} 81--95, 1984.

\bibitem{kn:Marg98} Margolus, Norman, ``Crystalline computation,'' In \cite{kn:Hey98} 267--305, 1999.

\bibitem{kn:Marg03} Margolus, Norman, ``Universal Cellular Automata Based on the Collisions of Soft Spheres,'' In \cite{kn:Ada01} 231--260, 2003.

\bibitem{kn:MM01} Mart\'{\i}nez, Genaro J. and McIntosh, Harold V., ``ATLAS: Collisions of gliders like phases of ether in rule 110,'' \url{http://uncomp.uwe.ac.uk/genaro/Papers/Papers_on_CA.html}, 2001.

\bibitem{kn:MMA} Mart\'{\i}nez, Genaro J., Morita, Kenichi, Adamatzky, Andrew, and Margenstern, Maurice, ``Logic Gates in Elementary Cellular Automata with Memory,'' {\em in elaboration}.

\bibitem{mershin_2004} Mershin~A., Kolomenski~A.~A., Schuessler~H.~A., Nanopoulos~D.~V., ``Tubulin dipole moment, dielectric constant and quantum behavior: computer simulations, experimental results and suggestions,'' {\em BioSystems} {\bf 77} 73--85, 2004. \url{arXiv:physics/0402053v1}  

\bibitem{kn:Mc90} McIntosh, Harold V. (1990) ``Wolfram's Class IV and a
Good Life,'' {\em Physica D} {\bf 45} 105--121.

\bibitem{kn:Mc09} McIntosh, Harold V. (2009) {\em One Dimensional
Cellular Automata}, Luniver Press.

\bibitem{mills_2008} Mills,~Jonathan~W., ``The nature of the Extended Analog Computer,'' {\em Physica D} {\bf 237} 1235--1256, 2008.
 
\bibitem{kn:MMI99} Morita, Kenichi, Margenstern, Maurice, and Imai, Katsunobu, ``Universality of reversible hexagonal cellular automata,'' {\em Theoret. Informatics Appl.} {\bf 33} 535--550, 1999.

\bibitem{kn:MMS06} Mart\'{\i}nez, Genaro J., McIntosh, Harold V., and Seck-Tuoh-Mora, J. C., ``Gliders in Rule 110,'' {\em Int. J. of Unconventional Computing} {\bf 2(1)} 1--49, 2006.

\bibitem{kn:MMS08} Mart\'{\i}nez, Genaro J., McIntosh, Harold V., Seck-Tuoh-Mora, Juan C., and Chapa-Vergara, Sergio V., ``Determining a regular language by glider-based structures called {\it phases f$_i$\_1} in Rule 110,'' {\em Journal of Cellular Automata} {\bf 3(3)} 231--270, 2008.

\bibitem{kn:MMS11} Mart\'{\i}nez, Genaro J., McIntosh, Harold V., Seck-Tuoh-Mora, Juan C., and Chapa-Vergara, Sergio V., ``Reproducing the cyclic tag system developed by Matthew Cook with Rule 110 using the phases f$_1$\_1,'' {\em Journal of Cellular Automata} {\bf 6(2-3)} 121--161, 2011.

\bibitem{kn:Mins67} Minsky, Marvin, {\em Computation: Finite and Infinite Machines}, Prentice Hall 1967.

\bibitem{kn:MNG04} Moreira, Andr\'es, Boccara, Nino, and Goles, Eric, ``On conservative and monotone one-dimensional cellular automata and their particle representation,'' {\em Theoretical Computer Science} {\bf 352(2)} 285--316, 2004.

\bibitem{kn:Mor08} Morita, Kenichi, ``Reversible computing and cellular automata---A survey'', {\em Theoretical Computer Science} {\bf 395} 101--131, 2008.

\bibitem{nakagaki_2000maze} Nakagaki T., Yamada H., and Toth A., ``Maze-solving by an amoeboid organism,'' {\em Nature} {\bf 407} 470, 2000.

\bibitem{kn:NW06} Neary, Turlough and Woods, Damien, ``P-completeness of cellular automaton Rule 110,'' {\em Lecture Notes in Computer Science} {\bf 4051} 132--143, 2006.

\bibitem{kn:Piva07} Pivato, Marcus, ``Defect particle kinematics in one-dimensional cellular automata,'' {\em Theoretical Computer Science} {\bf 377} 205--228, 2007.

\bibitem{rasmussen_1990} Rasmussen, S., Karampurwala, H., Vaidyanath, R., Jensen, K.S., and Hameroff, S., ``Computational connectionism within neurons: A model of cytoskeletal automata subserving neural networks,'' {\em Physica D} {\bf 42} 428--449, 1990.

\bibitem{kn:Red10} Redeker, Markus (2010) ``Flexible Time and the
Evolution of One-Dimensional Cellular Automata,'' {\em Journal of
Cellular Automata} {\bf 5(4-5)} 273--287.

\bibitem{kn:SBC07} Sapin, E., Bailleux, O., Chabrier, J. J., and
Collet, P.  (2004) ``A New Universal Cellular Automaton Discovered by
Evolutionary Algorithms,'' {\em Lecture Notes in Computer Science} {\bf
3102} 175--187.

\bibitem{kn:SCJ09} Lun Shi, Fangyue Chen, and Weifeng Ji, ``Gliders, Collisions and Chaos of Cellular Automata Rule 62,'' {\em International Workshop on Chaos-Fractals Theories and Applications}, 221--225, 2009.

\bibitem{kn:SHR05} Shalizi, Cosma R., Haslinger, Robert, Rouquier, Jean-Baptiste, Klinkner, Kristina L., and Moore, Cristopher, ``Automatic filters for the detection of coherent structure in spatiotemporal systems'', {\em Physical Review E} {\bf 73(3)} 036104, 2005.

\bibitem{stojanovic_2003} Stojanovic, M. N. and Stefanovic D. (2003) ``Deoxyribozyme-based half-adder,'' {\em J. Am. Chem. Soc.} {\bf 125} 6637.

\bibitem{kn:Toff98} Toffoli, Tommaso, ``Non-Conventional Computers'', In {\em Encyclopedia of Electrical and Electronics Engineering} {\bf 14} (John Webster Ed.), Wiley \& Sons, 455--471, 1998.

\bibitem{kn:Toff02} Toffoli, Tommaso, ``Symbol Super Colliders,'' In \cite{kn:Ada02}, 1--22, 2002.

\bibitem{kn:Wolf02} Wolfram, Stephen, {\em A New Kind of Science}, Wolfram Media, Inc., Champaign, Illinois 2002.

\bibitem{kn:WO08} Wolz, D. and de Oliveira, P.P.B. (2008) ``Very
effective evolutionary techniques for searching cellular automata rule
spaces,'' {\em Journal of Cellular Automata} {\bf 3(4)} 289--312.

\bibitem{kn:Wue99} Wuensche, Andrew, ``Classifying Cellular Automata Automatically,'' {\em Complexity} {\bf 4(3)} 47--66, 1999.

\end{thebibliography}
\end{document}